\documentclass[3p,times]{elsarticle}

\usepackage{ecrc}

\volume{00}
\firstpage{1}
\journalname{Sustainable Energy, Grids and Networks}

\runauth{E. Figini et al.}

\jid{Appl. Energy}

\jnltitlelogo{SEGAN}
\usepackage[figuresright]{rotating}
\usepackage[utf8]{inputenc}
\usepackage[T1]{fontenc}
\usepackage{siunitx}

\usepackage{algorithm}
\usepackage{algorithmicx}
\usepackage{algpseudocode}
\usepackage{courier}

\usepackage{amsmath,amsfonts,amssymb,amsthm}
\usepackage{mathtools}
\usepackage{mathrsfs}	
\usepackage{subdepth}	

\usepackage{array}	
\usepackage[inline]{enumitem} 

\PassOptionsToPackage{hyphens}{url}\usepackage{hyperref}

\usepackage[sort,compress,capitalise]{cleveref}	
\usepackage{url}

\usepackage{xcolor} 
\usepackage{xspace} 
\usepackage{lipsum}

\usepackage{tikz}
\usetikzlibrary{babel}  
\usepackage[straightvoltages,nooldvoltagedirection]{circuitikz}
\usetikzlibrary{shapes,arrows}
\usepackage{verbatim}
\usepackage{multirow}
\usepackage{svg}
\usepackage{booktabs}
\usepackage{subcaption}
\usepackage[intoc]{nomencl}
\makenomenclature

\newdefinition{rmk}{Remark}
\newcommand{\cs}[3]{#1_{\text{#2}}^{\text{#3}}}
\newcommand{\sref}[1]{(\ref{#1})}
\newcommand{\hl}[1]{\textcolor{black}{#1}} 

\begin{document}

\begin{frontmatter}


\dochead{}
\title{Achieving Dispatchability in Data Centers: Carbon and Cost-Aware Sizing of Energy Storage and Local Photovoltaic Generation}
\author{Enea Figini\corref{cor1}} 
\ead{enea.figini@epfl.ch}
\author{Mario Paolone}
\ead{mario.paolone@epfl.ch}
\address{Distributed Electrical Systems Laboratory (DESL) EPFL-STI-IEL-DESL ELL 116 (Bâtiment ELL) Station 11 CH-1015 Lausanne}
\cortext[cor1]{Corresponding author.}
\begin{abstract}
    \hl{Data centers are large electricity consumers due to the high consumption needs of servers and their cooling systems. With the rapid growth of crypto-currency and artificial intelligence, their electricity consumption is expected to increase substantially. With the electricity sector being responsible for a large share of global greenhouse gas (GHG) emissions, it is important to lower the carbon footprint of data centers to meet GHG emissions targets set by international agreements. Moreover, uncontrolled data center integration into power distribution grids increases the stochasticity of electricity demand, thus increasing the need for reserve capacity and leading to operational inefficiencies and higher emissions. \\
    This work provides a method to size a \emph{PhotoVoltaic} (PV) system and an \emph{Energy Storage System} (ESS) for an existing data center looking to reduce both its carbon footprint and demand stochasticity through day-ahead dispatching. A scenario-based optimization framework is developed to jointly size the PV and ESS, minimizing the expected operational and capital expenditures and the carbon footprint of the data center complex. The model considers the life cycle assessments (LCA) of the systems and the dynamic carbon intensity of the upstream electricity supply. Case studies in different Swiss cantons and regions of Germany emphasize the need for location-aware sizing processes since the obtained optimal solutions strongly depend on the local electricity carbon footprint and on the irradiance conditions. The maximum carbon footprint reduction reaches approximately 50\% in Germany and 4\% in Switzerland. Installed power generation and energy storage capacities vary by up to 36 times across regions.}
    \begin{center}
        \includegraphics[width=1\linewidth]{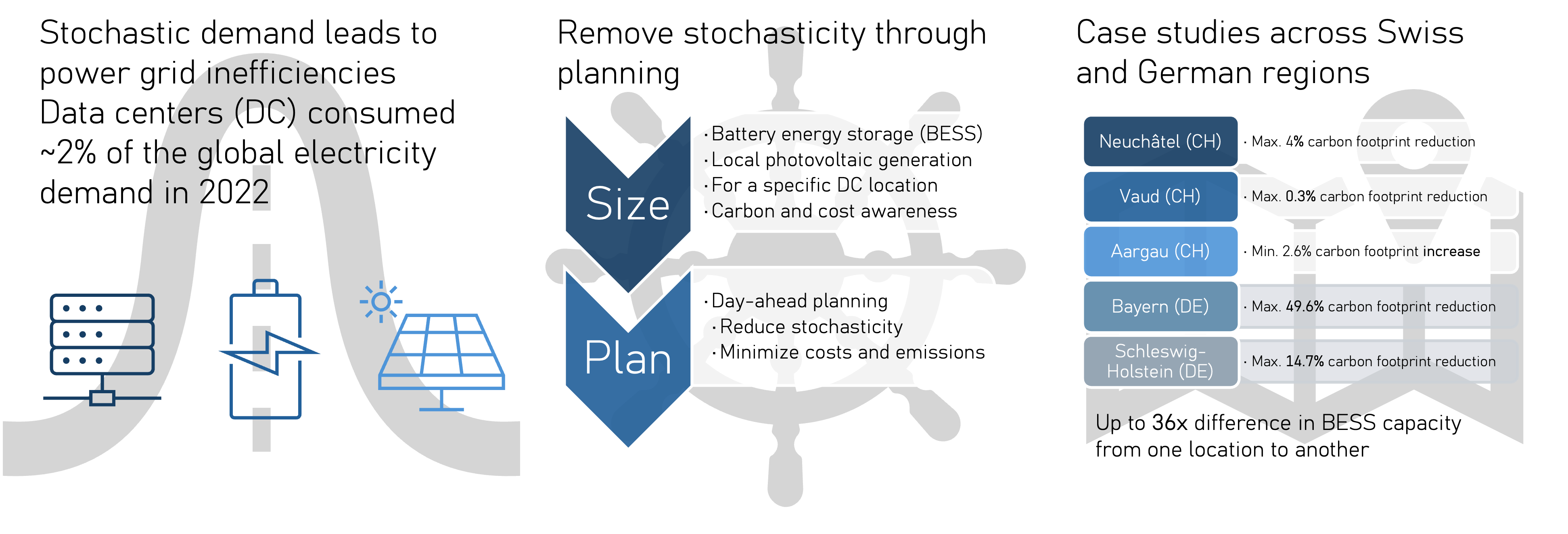}
    \end{center}
\end{abstract}
\end{frontmatter}


\section{Introduction}
\label{introduction}
Data centers consumed \SI{460}{\tera\watt\hour} of electricity worldwide in 2022, representing almost 2\% of the total electricity demand \cite{eren_cam_electricity_2024}. The \emph{International Energy Agency} (IEA) expects \hl{the global annual electricity demand from data centers} to range between 650 and \SI{1050}{\tera\watt\hour} in 2026, potentially accounting for up to 4.5\% of the global electricity demand. In Ireland and Denmark, the data center electricity consumption represents respectively 13\% and 8\% of the country's demand and it is expected to grow to 30\% and 20\%\hl{, respectively,} by 2026. With the electricity and heat sector accounting for 14.8 billions of tons of \SI{}{CO_2 eq} \cite{iea_electricity_2023} (i.e., approximately 25\% of global \emph{GreenHouse Gases} (GHG) emissions worldwide), reducing the carbon footprint of data centers is important to meet GHG emissions targets set by international agreements (e.g., \cite{noauthor_paris_2015}). Additionally, uncontrolled data centers tend to increase the stochasticity of the power system demand and to contribute to an environmentally and economically suboptimal operation of power grids \cite{bhagwat_effectiveness_2017, elcom_equilibre_2024}. For these reasons, carbon-aware solutions for data centers have been widely explored in the literature. \hl{An advanced search on IEEE \textit{Xplore} combining the keywords “green” and “data center” for the period 2020–2024 results in over 2,500 publications \cite{noauthor_green_nodate}, covering strategic planning/design, modeling, energy efficiency, green workload schedulers, waste heat recovery, etc. For example, in \cite{li_transforming_2020}, a cooling control algorithm using reinforcement learning is presented and studied using a simulation model. The algorithm trains two neural networks predicting the best next control action and the system state (in this case \emph{Power Usage Efficiency} PUE) and server room temperature. In \cite{yuan_biobjective_2021}, the authors present an optimization problem to schedule tasks in \emph{Distributed Green Data Centers} (DGDCs), considering local factors (s.a. electricity prices, wind speed, solar irradiance, etc). They maximize a biobjective function consisting of the aggregate revenue and \emph{Quality Of Service}\footnote{The quality of service quantifies the performance of a data center to serve the workloads/tasks of users.} (QoS) and use classified M/M/1 queuing models\footnote{\hl{An M/M/1 queue models the length of a queue in a single server system where job arrivals follow a Poisson process and execution times follow an exponential distribution. Such queue models are widely used to study workload scheduling performance for example.}} to model workload behaviour. As a final example, \cite{abdennadher_carbon_2022} showcases a framework to efficiently place data center locations to minimize the emissions of the power network under study, calculating the electricity dispatch of the system through a direct current optimal power flow. The paper shows that locating data centers close to renewable generation can be suboptimal and that grid operations and limits should be considered in the placement process.}\\
In practice, \emph{Data Center Operators} (DCO) often purchase renewable energy or renewable energy credits to match their annual demand \cite{noauthor_google_2019, noauthor_microsoft_2024}. This approach has been shown to be inefficient and misleading as it cannot guarantee an actual reduction of the carbon footprint of the entity making the purchase \cite{bjorn_renewable_2022, shannon__osaka_buying_2023}, thus highlighting the need for precise and granular carbon accounting \cite{noauthor_carbon_2023}. 
\subsection{Background}
\label{sec:previous work}
\hl{Recent literature highlights the relevance of integrating renewable energy systems and energy storage technologies to improve dispatchability and sustainability (e.g., \cite{luo_overview_2015, schmidt_monetizing_2023, emrani_comprehensive_2024}). Various energy storage technologies are studied in \cite{luo_overview_2015}, with recommendations of technologies for diverse applications, such as energy management, voltage regulation, load leveling, etc. A recent and comprehensive review of hybrid energy storage systems integrated with renewables is provided in \cite{emrani_comprehensive_2024}. The study compares a wide range of storage technologies including batteries, thermal storage, compressed air, flywheels, and hydrogen storage. The use of these technologies is assessed in terms of cost, efficiency, maturity, environmental friendliness, etc. Moreover, it reviews optimization approaches for sizing and operation, including stochastic multi-objective optimization and techno-economic models. Notably, battery energy storage systems are highlighted as particularly well-suited for high-power, high-energy-demand applications, such as the ones associated to modern data center operations. Building on this, methods for power system optimization under uncertainty are reviewed in \cite{roald_power_2023} and stochastic optimization is generally identified as crucial in capturing the uncertainty in integrated energy systems planning and operation \cite{wu_enhanced_2024}. These approaches are particularly relevant for data centers, which increasingly rely on the integration of renewable generation and require flexibility to achieve reliability objectives under uncertain conditions.}\\
Strategic planning of data centers has been studied with a general focus on methods to reduce their operational costs. For instance, \cite{guo_integrated_2021} formulates a multi-objective optimization problem to determine the optimal capacity and location of a data center and  associated energy storage systems, minimizing the costs and maximizing the QoS. \cite{yong_scheduling_2024} proposes a day-ahead scheduling of data centers based on the concept of a \emph{Virtual Power Plant} (VPP), leveraging the flexibility of workloads (i.e., using workload migration and shifting\footnote{Workload migration refers to the process of moving workloads from one computing environment to another, while workload shifting refers to moving workloads in time.}), and the flexibility of back-up power devices (i.e., UPSs\footnote{\emph{Uninterruptible Power Supplies} (UPS) are back-up power systems used to mitigate the risks on the data center supply unavailability due to temporary black-outs of the power grid.}). Some works on the planning of data centers consider their direct carbon footprint. For instance, \cite{kong_greenplanning_2016} formulates a framework to select energy resources, balancing energy sources with grid power and storage in terms of cost, direct emissions and service availability. \cite{thompson_optimization_2016} presents a sizing method for \emph{battery energy storage systems} (BESS) in data center microgrids. The method allows for costs and emissions reductions by leveraging batteries to shift the peak period demand to off-peak periods. In \cite{wang_day-ahead_2017}, a day-ahead direct emission aware-planning of data centers is formulated, where conventional power units and energy storage systems are jointly optimized with batch workload\footnote{Batch workloads are automatically completed pre-defined jobs/tasks.} allocation. In \cite{rahmani_modelling_2020}, the sizing of wind and PV generation to minimize the operational cost and emissions of a green data center is addressed. 
\subsection{Contributions and proposed method}
As discussed in \cref{sec:previous work}, although some studies have been focusing on carbon-aware strategic planning of data centers, the majority of the available literature focuses on operational cost minimization. The studies that take the carbon emissions into consideration rely on simplistic approaches by focusing on direct emissions \cite{kong_greenplanning_2016, wang_day-ahead_2017}, or on qualitative information \cite{rahmani_modelling_2020}. In \cite{dou_carbon-aware_2017, radovanovic_carbon-aware_2021, lin_adapting_2023} the hourly average carbon emissions of the grid are taken into account to reduce the footprint of data centers. \cite{lin_adapting_2023} highlights the need for a day-ahead scheduling of the data center power consumption to effectively reduce the carbon emissions. These studies leverage the intrinsic flexibility of the data center, but do not study the flexibility of co-located \emph{Distributed Energy Resources} (DER) and, therefore, do not assess the optimal sizing of these DER associated to data centers. Finally, while \cite{acun_carbon_2022} proposes a framework to design carbon-aware data centers with DER, it does not consider economic costs and focuses on 24/7 renewable coverage for the data center, which might not be the optimal system design in terms of total carbon footprint since the \SI{}{CO_2} content of the data center electricity supply varies over time and is location dependent.\\
In view of the above, this work addresses the problem of carbon-aware sizing of energy storage and photovoltaic generation for existing data centers targeting day-ahead dispatchability (i.e., to track, in real-time, a power profile computed day-ahead). The sizing method allows for geographically and temporally granular grid carbon emissions awareness, while also enabling custom operational cost considerations. The proposed method relies on scenario-based optimization, and is thus capable of taking into account the non-parametric forecasts of the grid carbon intensity, electricity costs, solar irradiance and data center demand. Local considerations (e.g., demand specificities, electricity pricing schemes, etc.) can be easily included in the framework by using precise local data and custom forecasting methods for the stochastic variables\footnote{The forecasting methods themselves are not the focal point of this work. Nevertheless, they significantly impact the results of the framework.}. Since this work focuses on day-ahead dispatchability, the focus is placed on short-term energy storage solutions (such as BESS).
\section{Method}
\label{method}
\subsection{Overview}
\subsubsection{Problem statement}
\begin{figure}
    \centering
    \includegraphics[width=.5\textwidth]{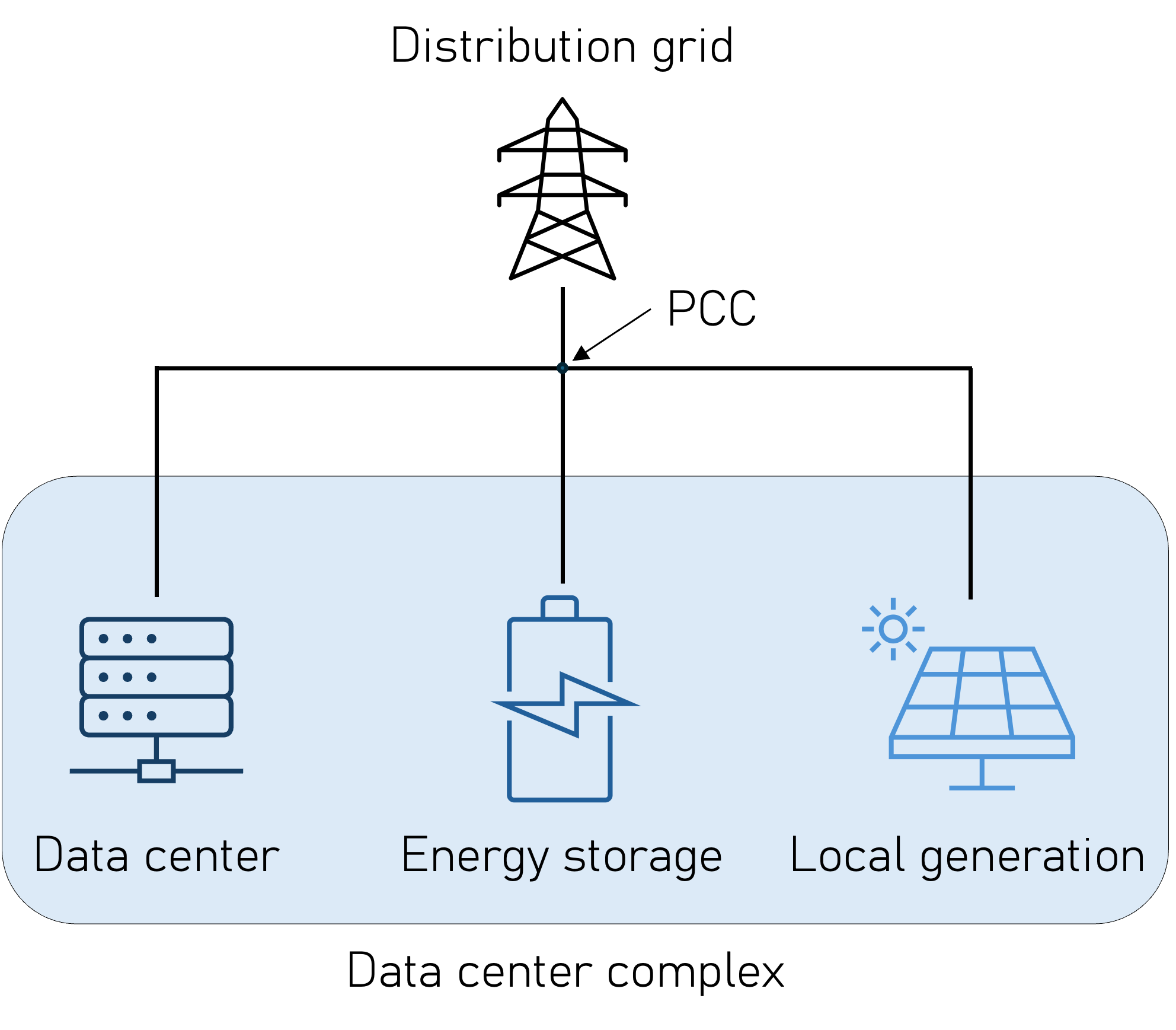}
    \caption{Schematic view of the resources configuration.}
    \label{fig:diagram}
\end{figure}

The method addresses the optimal sizing of local energy storage and photovoltaic generation to achieve the dispatchability of a data center. The problem determines the optimal ratings of an \emph{Energy Storage System} (ESS) and PV generation to minimize the expected value of a multi-objective function consisting in the weighted sum of carbon and financial costs. The configuration that is considered is shown schematically in \cref{fig:diagram}.
The method allows to consider the stochasticity of the data center demand, the solar irradiance, the carbon intensity of the imported electricity and the electricity tariffs, optimizing the system's expected value of the carbon and financial cost over $\cs{N}{tp}{}$ typical days with $\cs{N}{sc}{}$ daily scenarios (for each stochastic quantity mentioned above).
\subsubsection{Working hypotheses}
\label{sec:hypotheses}
The approach works with the following hypotheses:
\begin{itemize}
    \item The data center complex (i.e., the data center and the associated DERs) is owned by a single entity: the DCO. This ensures that the resources can share common objectives and allows for simple aggregation of investment and operational costs.
    \item \hl{The curtailment of the PV system is neglected. Since the focus of the paper is on the planning phase, instances of PV curtailment are expected to be rare due to the co-optimization of PV capacity with energy storage. Furthermore, this hypothesis reduces the complexity of the model.}
    \item The rated power at the \emph{Point of Common Coupling} (PCC) is given (i.e., the DCO has an agreement in place with the power Distribution System Operator regarding the rating of the data center electricity supply). \hl{The agreement includes a pre-defined day-ahead consumption tracking accuracy, later referred to as $\cs{\epsilon}{t}{}$.} 
    \item The efficiency of the ESS is assumed to be power-independent and constant. \hl{This assumption is justified in planning problems since the battery undergoes periodic cycles and using the round-trip efficiency does not lead to incorrect results. More sophisticated methods (such as those in \cite{ollas_battery_2023}) should be considered in operational/control problems.}
\end{itemize}
\subsection{Problem formulation}
A stochastic multi-objective convex minimisation problem with operational constraints is proposed. The problem determines the optimal ratings of the ESS and PV generation to minimize the expected value of an aggregated objective function, while ensuring the dispatchability of the data center complex. The objective function consists in the weighted sum of two objectives: the first aims at minimizing the carbon costs of the system (in \SI{}{\gram CO_2eq}), while the latter targets the financial costs (in ¤, the generic currency symbol). \cref{tab:objectives} summarizes the aspects that are considered in the objective function. 
\begin{table}[h]
    \centering
    \begin{tabular}{c|c}
         \textbf{Carbon costs in \SI{}{\g CO_2eq}} & \textbf{Financial costs in ¤}\\ \hline
         The emissions from grid electricity imports & Electricity bill \\
         ESS equivalent emissions & ESS investment, operation and maintenance \\
         Generation equivalent emissions & Generation investment, operation and maintenance
    \end{tabular}
    \caption{Components of the multi-objective function.}
    \label{tab:objectives}
\end{table}
The optimization process is performed under multiple operational constraints on the ESS dynamics, aging, efficiency and ratings, on the PV generation aging and ratings, as well as on the grid connection capacity. The problem aims at minimizing the expected value of the objective function, over $\cs{N}{tp}{}$ time horizons of $W$ hours each, and over $\cs{N}{sc}{}$ scenarios per time horizon. For every time horizon, the problem computes an optimal dispatch plan (i.e., $\cs{P}{pcc}{}(t)$, for $t\in[0, W]$) that is feasible for all the scenarios generated for that particular horizon. In other words, the sizing computes a dispatch plan for $\cs{N}{tp}{}$ $W$-long time horizons. In this paper, a $W=\SI{24}{\hour}$ time horizon is used, since it is a typical dispatching use-case due to the structure of the electricity market \cite{sossan_achieving_2016, gupta_coordinated_2022}.

\subsection{Mathematical model}
\subsubsection{Local glossary}
The parameters used in the optimization problem are listed in this section. Note that $N$ is the number of discrete time steps in the time horizon $W$ (i.e., the number of steps in a typical day) and $\Delta T$ is the duration of a time step (thus, $W = N\cdot \Delta T$ is the length of the horizon in hours). $M$ is the product of $\cs{N}{tp}{}$ and $\cs{N}{sc}{}$; it corresponds to the total number of scenarios considered in the sizing problem. Matrices are highlighted in bold and, for a given matrix $\cs{\textbf{M}}{}{}$, the entry at row $i$ and column $j$ is referred to as $\cs{\textbf{M}}{}{}[i][j]$. The decision variables are the power profile of the storage system ($\cs{\textbf{P}}{ess}{} \in \mathbb{R}^{N\times M}$), the rated capacity of the storage system ($\cs{E}{ess}{rated} \in \mathbb{R}$) and the rated power of the PV generation system ($\cs{P}{gen}{rated} \in \mathbb{R}$). To enhance readability, the other variables are only listed with their nature and dimensions. If their meaning is not clear, the reader can refer to the glossary in \ref{nomenclature}.

\begin{itemize}
    \item Auxiliary variables:
    \begin{itemize}
        \item $\cs{\textbf{E}}{ess}{} \in \mathbb{R}^{(N+1)\times M}$
        \item $\cs{\textbf{P}}{pcc}{}$, $\cs{\textbf{P}}{gen}{}$, $\cs{\textbf{P}}{ess}{}$,
              $\cs{\textbf{P}}{ess}{conv}$, $\cs{\textbf{P}}{ess}{charge}$,
              $\cs{\textbf{P}}{ess}{discharge}$, 
              $\cs{\textbf{P}}{pcc}{load}$, $\cs{\textbf{P}}{pcc}{gen}$,
              $\cs{\textbf{z}}{ess}{relaxed}$, $\cs{\textbf{z}}{pcc}{relaxed}$ $\in \mathbb{R}^{N\times M}$
        \item $\cs{\textbf{z}}{ess}{}$, $\cs{\textbf{z}}{pcc}{}$ $\in \mathbb{Z}_2^{N\times M}$, with $\mathbb{Z}_2 = \{0, 1\}$
        \item $\cs{\textbf{P}}{pcc}{dispatch} \in \mathbb{R}^{N\times \cs{N}{tp}{}}$
        \item $\cs{\textbf{P}}{pcc}{max} \in \mathbb{R}^{1\times M}$
        \item $\cs{P}{ess}{rated}$, $\cs{E}{ess}{rated}$, $\cs{P}{gen}{rated}$
              $\cs{C}{e}{pcc}$, $\cs{C}{e}{ess}$, $\cs{C}{e}{gen}$, $\cs{c}{ess}{}$, $\cs{c}{gen}{}$, $\cs{c}{el}{energy}$, $\cs{c}{el}{power} \in \mathbb{R}$
    \end{itemize}
        
    \item Input parameters:
    \begin{itemize}
        \item $\cs{\textbf{P}}{load}{}$, $\cs{\textbf{i}}{ghi}{}$, 
                $\cs{\textbf{C}}{i}{pcc}$, $\cs{\textbf{p}}{el}{cons}$, $\cs{\textbf{p}}{el}{inj} \in \mathbb{R}^{N\times M}$
        \item $\cs{N}{}{}$, $\cs{M}{}{}$, $\cs{N}{tp}{}$, $\cs{N}{sc}{}$, $\Delta T$, $W$, $\cs{W}{days}{}$, $w$, 
               $\cs{P}{pcc}{rated}$, $\cs{C}{i}{ess}$, $\cs{C}{e, LCA}{ess}$, $\cs{C}{e, LCA}{gen}$,
               $\cs{r}{ess}{p2e}$, $\cs{r}{gen}{ghi2p}$, $\cs{i}{ghi}{max}$,
               $\cs{\textit{SoC}}{ess}{min}$, $\cs{\textit{SoC}}{ess}{max}$, $\cs{\textit{SoC}}{ess}{start}$,
               $\cs{L}{ess}{cycles}$, $\cs{L}{ess}{calendar}$, $\cs{L}{gen}{calendar}$, $\cs{E}{ess}{start}$, 
               $\cs{E}{ess}{min}$, $\cs{E}{ess}{max}$, $\cs{P}{pcc}{rated}$, $\cs{P}{ess}{rated, max}$, $\cs{a}{ess}{}$, $\cs{c}{ess}{life}$,
               $\cs{c}{ess}{energy}$, $\cs{c}{ess}{power}$, $\cs{c}{gen}{power}$, $\cs{p}{el}{power}$, $\cs{\eta}{ess}{}$, $\cs{\epsilon}{t}{}$, $\cs{M}{ess}{}$, $\cs{M}{pcc}{} \in \mathbb{R}$
    \end{itemize}
\end{itemize}
Also, note that in the following, inequalities between a matrix and a scalar (e.g., $\textbf{V}\geq s$) mean that all entries of the matrix must satisfy the inequality (e.g., all entries of $\textbf{V}$ are larger or equal to $s$). 
\subsubsection{Formulation}
\label{Formulation}
The proposed objective function $\cs{F}{obj}{}$ is given in \sref{eq:objective function}, where the symbols \hl{$\cs{C}{x}{y}$ represent the carbon costs and the symbols $\cs{c}{x}{y}$ represent the financial costs. The weight $w$ in \SI{}{\g CO_2eq \per}¤ ponders the two objectives.}

\begin{equation}
    \label{eq:objective function}
    \begin{aligned}
        \cs{F}{obj}{}\left(\cs{\textbf{P}}{ess}{}, \cs{E}{ess}{rated}, \cs{P}{gen}{rated}\right)= \cs{C}{e}{pcc} + \cs{C}{e}{ess} + \cs{C}{e}{gen} + w\cdot\left(\cs{c}{ess}{} + \cs{c}{gen}{} + \cs{c}{el}{energy} + \cs{c}{el}{power}\right)
    \end{aligned}
\end{equation}
In \sref{eq:set o}, $\mathcal{O}$ is used to denote the set of operational constraints (i.e., the physical constraints considered in the sizing problem). For instance, \sref{const:rated ess power} limits the ESS power to its rated values, while the energy contained in the ESS is constrained by \sref{const:eess start}, which forces a predefined energy to be stored at the start of every day in the optimization time window. \sref{const:ess energy} sets the lower and upper bounds for the energy stored in the system. Constraint \sref{const:ess dynamics} links the energy stored in the ESS with its power. The power limit at the point of common coupling is set through \sref{const:ppcc lims} and the power at the PCC is computed in \sref{const:kirchoff pcc} (note that the assumption that losses can be neglected is made). Finally, \sref{const:prated gen} ensures that the rated power of the generation is a positive number.
\begin{subequations}
\label{eq:set o}
\begin{alignat}{1}
\mathcal{O} = [ & -\cs{P}{ess}{rated} \leq \cs{\textbf{P}}{ess}{conv} \leq \cs{P}{ess}{rated}, \label{const:rated ess power}\\
                & \cs{\textbf{E}}{ess}{}[0][j] = \cs{E}{ess}{start},~\text{for } [j \in \mathbb{N}: 0\leq j \leq M-1], \label{const:eess start}\\
                & \cs{E}{ess}{min} \leq \cs{\textbf{E}}{ess}{} \leq \cs{E}{ess}{max}, \label{const:ess energy}\\
                & \cs{\textbf{E}}{ess}{}[k+1][j] = \cs{\textbf{E}}{ess}{}[k][j]+\Delta T\cdot\cs{\textbf{P}}{ess}{}[k][j], ~\text{for } [k \in \mathbb{N}: 0\leq k \leq N-1], ~\text{and } [j \in \mathbb{N}: 0\leq j \leq M-1], \label{const:ess dynamics} \\
                & -\cs{P}{pcc}{rated} \leq \cs{\textbf{P}}{pcc}{} \leq \cs{P}{pcc}{rated}, \label{const:ppcc lims}\\
                & \cs{\textbf{P}}{pcc}{} = \cs{\textbf{P}}{ess}{conv} + \cs{\textbf{P}}{load}{} - \cs{\textbf{P}}{gen}{}, \label{const:kirchoff pcc}\\
                & \cs{P}{gen}{rated} \geq 0 \label{const:prated gen}]
\end{alignat}
\end{subequations}
In \sref{eq:set a}, $\mathcal{A}$ is the set of auxiliary constraints (i.e., the artificial constraints used to formulate the problem). Constraints \sref{const:soc min}, \sref{const:soc max} and \sref{const:soc start} link the user defined parameters $\cs{\textit{SoC}}{ess}{min/max/start}$ and their equivalent stored energies, which are then used in constraints \sref{const:eess start} and \sref{const:ess energy}. \hl{For all typical time horizons, constraint \sref{const: ess as buffer} ensures that the ESS does not get charged nor discharged on average}\footnote{This constraint might not be needed in a day-to-day operation but is relevant in a planning phase, as the considered time horizons are not necessarily contiguous.}. \hl{Moreover, \sref{const:pcc split} to \sref{const:indicator pcc gen} are used to recover the generation and consumption components of the PCC dispatch. Constraint \sref{const:pcc split} ensures that the sum of those components is equal to the PCC power while \sref{const:pcc load g0}, \sref{const:pcc gen l0}, \sref{const:indicator pcc load} and \sref{const:indicator pcc gen} ensure that at least one of $\cs{\textbf{P}}{pcc}{load}$ and  $\cs{\textbf{P}}{pcc}{gen}$ is equal to zero at any timestep of any scenario.} Similarly, \sref{const: pess split} to \sref{const:indicator ess discharge} enable the splitting of the ESS power. Note the user input $\cs{P}{ess}{rated, max}$ which defines the maximum ESS power rating that is considered in the sizing problem (i.e., $|\cs{P}{ess}{rated}| \leq \cs{P}{ess}{rated, max}$). More detail on the splitting of $\cs{\textbf{P}}{pcc}{dispatch}$ (and $\cs{\textbf{P}}{ess}{}$) is provided in Remark\xspace\ref{rmk:Ppcc splitting}. Constraint \sref{const:ess efficiency} links the storage power and the converter power through the efficiency of the system $\cs{\eta}{ess}{}$. \hl{The constraint in \sref{const:ess p2e} forces a dependency between the energy and power ratings of the ESS through the user-defined parameter $\cs{r}{ess}{p2e}$, which is used to make the problem convex (in particular, it simplifies \sref{const:cost ess} as detailed further below).} Constraint \sref{const: dispatch} ensures that the PCC power for every scenario of a given typical day is equal to the dispatch of that day (note that the transformation matrix $\cs{\textbf{T}}{m}{} \in \mathbb{Z}_2^{\cs{N}{tp}{}\times M}$ has to be built appropriately). Constraints \sref{const: pcc max} and \sref{const: pcc min} recover the maximum power consumption and generation of every typical day, so that they can be billed appropriately. Finally, the photovoltaic power is defined in \sref{const:pgen}. Following \cite{richardson_validation_2019}, the power generation is modeled as a linear relationship between the solar irradiance [\SI{}{\watt\per\meter^2}] and the plant peak power rating (defined as $\cs{i}{ghi}{max}$ \SI{}{\watt\per\meter^2}). Note that $\cs{r}{gen}{ghi2p}\in[0,\xspace 1]$ is a user-defined constant that can be used to adapt the slope of the linear relationship. 
\begin{subequations}
\label{eq:set a}
\begin{alignat}{1}
\mathcal{A} = [ & \cs{\textit{SoC}}{ess}{min}\cdot\cs{E}{ess}{rated} = \cs{E}{ess}{min}, \label{const:soc min}\\
                & \cs{\textit{SoC}}{ess}{max}\cdot\cs{E}{ess}{rated} = \cs{E}{ess}{max}, \label{const:soc max}\\
                & \cs{\textit{SoC}}{ess}{start}\cdot\cs{E}{ess}{rated} = \cs{E}{ess}{start}, \label{const:soc start}\\
                & \sum_{k=0}^{N-1}\left(\cs{\textbf{P}}{ess}{}\cdot\textbf{T}_\text{m}^\intercal\right)[k]=0, \label{const: ess as buffer}\\
                & \cs{\textbf{P}}{pcc}{} = \cs{\textbf{P}}{pcc}{load}+\cs{\textbf{P}}{pcc}{gen}, \label{const:pcc split}\\
                & \cs{\textbf{P}}{pcc}{load}\geq0, \label{const:pcc load g0}\\
                & \cs{\textbf{P}}{pcc}{gen}\leq0, \label{const:pcc gen l0}\\
                & \cs{\textbf{P}}{pcc}{load}\leq \cs{P}{pcc}{rated}\cdot(1-\cs{\textbf{z}}{pcc}{}), \label{const:indicator pcc load}\\
                & -\cs{\textbf{P}}{pcc}{gen}\leq \cs{P}{pcc}{rated}\cdot\cs{\textbf{z}}{pcc}{}, \label{const:indicator pcc gen}\\
                & \cs{\textbf{P}}{ess}{} = \cs{\textbf{P}}{ess}{charge} + \cs{\textbf{P}}{ess}{discharge}, \label{const: pess split}\\
                & \cs{\textbf{P}}{ess}{charge} \geq 0, \label{const:pess charge g0}\\
                & \cs{\textbf{P}}{ess}{discharge} \leq 0, \label{const:pess discharge l0}\\
                & \cs{\textbf{P}}{ess}{charge}\leq \cs{P}{ess}{rated, max}\cdot(1-\cs{\textbf{z}}{ess}{}), \label{const:indicator ess charge}\\
                & -\cs{\textbf{P}}{ess}{discharge}\leq \cs{P}{ess}{rated, max}\cdot\cs{\textbf{z}}{ess}{}, \label{const:indicator ess discharge}\\
                & \cs{\textbf{P}}{ess}{conv} = \cs{\textbf{P}}{ess}{charge}/\cs{\eta}{ess}{} + \cs{\eta}{ess}{}\cdot\cs{\textbf{P}}{ess}{discharge}, \label{const:ess efficiency}\\
                & \cs{P}{ess}{rated}=\cs{r}{ess}{p2e}\cdot\cs{E}{ess}{rated}, \label{const:ess p2e}\\
                & \cs{\textbf{P}}{pcc}{} = \cs{\textbf{P}}{pcc}{dispatch}\cdot\cs{\textbf{T}}{m}{}, \label{const: dispatch}\\
                & \textbf{1}^{N\times1} \cdot \cs{\textbf{P}}{pcc}{max} \geq \cs{\textbf{P}}{pcc}{load}, \label{const: pcc max} \\
                & \textbf{1}^{N\times1} \cdot \cs{\textbf{P}}{pcc}{max} \geq -\cs{\textbf{P}}{pcc}{gen}, \label{const: pcc min}\\
                & \cs{\textbf{P}}{gen}{} = \frac{\cs{r}{gen}{ghi2p}\cs{P}{gen}{rated}}{\cs{i}{ghi}{max}}\cdot\cs{\textbf{i}}{ghi}{} \label{const:pgen}]     
\end{alignat}
\end{subequations}

$\mathcal{F}$, in \sref{eq:set f}, gives the set of constraints used to compute the expected daily costs of the system. In the set, $j\in[j:0\leq j <M]$ is the index over typical days and scenarios, while $k\in[k:0\leq k <N]$ is the index over the steps in typical days. Constraints \sref{const:carbon ess}, \sref{const:carbon gen} and \sref{const:carbon pcc} model the expected carbon costs of the ESS, the generation and the PCC respectively.
\begin{subequations}
\label{eq:set f}
    \begin{alignat}{1}
    \mathcal{F} = [& \cs{C}{e}{ess} = \frac{\Delta T \cdot \cs{C}{i}{ess}}{M} \cdot \sum_{j=0}^{M-1} \sum_{k=0}^{N-1} \left|\cs{\textbf{P}}{ess}{}[k][j]\right|+\cs{E}{ess}{rated} \frac{W \cdot \cs{C}{e, LCA}{ess}}{\cs{L}{ess}{calendar}}, \label{const:carbon ess}\\
                   & \cs{C}{e}{gen} = \frac{W}{\cs{L}{gen}{calendar}} \cdot \cs{P}{rated}{gen} \cdot \cs{C}{e, LCA}{gen}, \label{const:carbon gen}\\
                   & \cs{C}{e}{pcc} = \frac{\Delta T}{M} \cdot \sum_{j=0}^{M-1} \sum_{k=0}^{N-1} \cs{\textbf{C}}{i}{pcc}[k][j] \cdot \cs{\textbf{P}}{pcc}{load}[k][j], \label{const:carbon pcc}\\
                   & \cs{c}{ess}{} = \left(\frac{\cs{E}{ess}{rated} \cdot W}{\cs{L}{ess}{calendar}} + \frac{\Delta T}{2 \cdot \cs{L}{ess}{cycles} \cdot M}\sum_{j=0}^{M-1}\sum_{k=0}^{N-1}\left|\cs{\textbf{P}}{ess}{}[k][j]\right|\right) \cdot \left(\cs{c}{ess}{energy}+\cs{c}{ess}{power} \cdot \cs{r}{ess}{p2e}\right), \label{const:cost ess}\\
                   & \cs{c}{gen}{} = \frac{W}{\cs{L}{gen}{calendar}} \cdot \cs{P}{gen}{rated} \cdot \cs{c}{gen}{power},\label{const:cost gen} \\
                   & \cs{c}{el}{energy} = \frac{\Delta T}{M} \cdot \sum_{j=0}^{M-1}\sum_{k=0}^{N-1}\cs{\textbf{p}}{el}{cons}[k][j] \cdot \cs{\textbf{P}}{pcc}{load}[k][j]+\cs{\textbf{p}}{el}{inj}[k][j] \cdot \cs{\textbf{P}}{pcc}{gen}[k][j], \label{const:cost energy}\\
                   & \cs{c}{el}{power} = \frac{1}{M}\sum_{j=0}^{M-1}\cs{p}{el}{power} \cdot \cs{\textbf{P}}{pcc}{max}[j] \label{const:cost power}\cdot\cs{W}{days}{}]
    \end{alignat}
\end{subequations}
\hl{In \sref{const:carbon ess}, equivalent ESS emissions are estimated. Within the scope of \emph{Life Cycle Assessments} (LCA) for energy storage systems, $\cs{C}{e, LCA}{ess}$ represents the \SI{}{gCO_2eq} emitted to manufacture and install a single kilowatt-hour of ESS. Over a given period, the emissions resulting from the asset are estimated by scaling $\cs{C}{e, LCA}{ess}$ according to the amount of aging that occurs in that period. The aging process is modeled as the superposition of calendar aging and cycling-based degradations \cite{schmalstieg_holistic_2014}. Emissions related specifically to calendar aging of the ESS are modeled in \cref{eq:calendar carbon ess}, which assumes a linear relationship between time and calendar degradation.}
\begin{equation}
\label{eq:calendar carbon ess}
    \cs{C}{e, cal}{ess} =\frac{W}{\cs{L}{ess}{calendar}}\cdot\cs{E}{ess}{rated}\cdot\cs{C}{e, LCA}{ess}
\end{equation}
\hl{To further account for the effects of cycling-based aging (e.g., for battery storage systems), the term $\cs{C}{i}{ess}$ \sref{eq:ci ess} is introduced, which quantifies the carbon emissions per \SI{}{\kWh} of energy throughput (note that $2\cdot\cs{L}{ess}{cycles}$ corresponds to the maximum throughput over the lifetime of a \SI{}{\kWh} of storage). Following  \cite{schmalstieg_holistic_2014}, cycling-based aging is assumed to be linearly growing with the ESS energy throughput $\cs{\textbf{E}}{throughput}{ess}$ \sref{eq:energy throughput}, which leads to \sref{eq:cycling-based emissions}. The total ESS-related emissions \sref{const:carbon ess} are computed as the sum of \sref{eq:calendar carbon ess} and \sref{eq:cycling-based emissions}.}
\begin{subequations}
\begin{alignat}{1}
    \cs{C}{i}{ess} &= \frac{\cs{C}{e, LCA}{ess}}{2\cdot\cs{L}{ess}{cycles}} \label{eq:ci ess}\\
    \cs{\textbf{E}}{throughput}{ess}[j] &= \Delta T\cdot\sum_{k=0}^{N-1}\left|\cs{\textbf{P}}{ess}{}[k][j]\right| \text{, for } 0\leq j<M \label{eq:energy throughput}\\
    \cs{C}{e, cycling}{ess} &= \frac{\cs{C}{i}{ess}}{M} \sum_{j=0}^{M-1}\cs{\textbf{E}}{throughput}{ess}[j] \label{eq:cycling-based emissions}
\end{alignat}
\end{subequations}
Similarly to \sref{eq:calendar carbon ess}, the emissions of the generation asset are estimated in \sref{const:carbon gen} using its life cycle assessment. $\cs{C}{e, LCA}{gen}$, which estimates the \SI{}{gCO_2eq} emitted to install a kilowatt of photovoltaic generation peak power. The carbon emissions related to the grid imports (i.e., $\cs{\textbf{P}}{pcc}{load}$) are computed in \sref{const:carbon pcc} using the carbon intensity of the imported electricity $\cs{\textbf{C}}{i}{pcc}$ in \SI{}{\gram CO_2eq\per\kWh}. \\
Finally, \sref{const:cost ess} to \sref{const:cost power} model the expected financial costs of the system over a given period. \hl{The cost attributed to the ESS aging is expressed in \sref{const:cost ess}. Similarly to the carbon emission modelling in \sref{const:carbon ess}, this cost is calculated by scaling the total installation cost, $\cs{c}{ess}{life}$ \sref{eq:c ess life}, by the average aging (i.e., $\cs{a}{ess}{}$) of the system during the optimization time horizon \sref{eq:aging}. However, as shown in \sref{eq:non-convexity}, the scaling results in a non-convexity caused by a division of two decision variables $\cs{P}{ess}{rated}$ and $\cs{E}{ess}{rated}$. To address this, the input parameter $\cs{r}{ess}{p2e}$ is introduced in \sref{eq:rp2e} to enforce a fixed ESS power-to-energy ratio (i.e., the ratio between these decision variables), allowing the formulation of a restrictive but convex constraint. As a result, \sref{const:cost ess} is reformulated as the product $\cs{c}{ess}{life} \cdot \cs{a}{ess}{}$, where $\cs{P}{ess}{rated}$ is replaced by $\cs{r}{ess}{p2e} \cdot \cs{E}{ess}{rated}$. A discussion of the practical implications of fixing $\cs{r}{ess}{p2e}$ is provided in Remark~\ref{rmk:p2e ratio} since this ratio is usually fixed by BESS manufacturers.}
\begin{subequations}
\begin{alignat}{1}
    \cs{c}{ess}{life}&=\cs{c}{ess}{energy}\cdot\cs{E}{ess}{rated} + \cs{c}{ess}{power}\cdot\cs{P}{ess}{rated} \label{eq:c ess life}\\
    \cs{a}{ess}{}&=W/\cs{L}{ess}{calendar}+\frac{\sum_{j=0}^{M-1}\cs{\textbf{E}}{throughput}{ess}[j]}{2\cdot\cs{E}{ess}{rated}\cdot\cs{L}{ess}{cycles}\cdot M} \label{eq:aging}\\
    \hl{\cs{a}{ess}{}\cs{c}{ess}{life}} &\hl{=}~\hl{... + \frac{\sum_{j=0}^{M-1}\cs{\textbf{E}}{throughput}{ess}[j]}{2\cdot\cs{E}{ess}{rated}\cdot\cs{L}{ess}{cycles}\cdot M} \cdot \cs{c}{ess}{power}\cdot \cs{P}{ess}{rated}} \label{eq:non-convexity}\\
    \cs{r}{ess}{p2e}&=\frac{\cs{P}{ess}{rated}}{\cs{E}{ess}{rated}} \label{eq:rp2e}
\end{alignat}
\label{eq:ess-cost-steps}
\end{subequations}
    
The financial cost of the generation asset is considered in \sref{const:cost gen}, which is identical to \sref{const:carbon gen} but in financial terms. Lastly, the electricity bill is estimated in \sref{const:cost energy} and \sref{const:cost power}. The first models the cost related to the energy consumption and production at the PCC (with prices per \SI{}{\kWh} set through $\cs{\textbf{p}}{el}{cons}$ and $\cs{\textbf{p}}{el}{inj}$), while the second models the cost related to the maximum \SI{15}{\minute} average power consumption. Although the maximum power is typically billed monthly or yearly, the worst case scenario over the optimization time window is used (which is equivalent to assuming one power bill per dispatch). Note that $\cs{p}{el}{power}$ likely needs to be recomputed (e.g., if the consumption site billing scheme bills $\cs{p}{el}{power, month}$ for each \SI{}{\kW} of monthly peak power, then $\cs{p}{el}{power}=\frac{12}{365}\cs{p}{el}{power, month}$).  
Combining \cref{eq:objective function}, \cref{eq:set o}, \cref{eq:set a}, and \cref{eq:set f}, the sizing optimization problem is formulated in \cref{eq:opti prob}.
\begin{equation}
\label{eq:opti prob}
\begin{aligned}
\min_{\cs{E}{ess}{rated},~\cs{P}{gen}{rated},~\cs{\textbf{P}}{ess}{}} \quad & \cs{F}{obj}{}\\
\textrm{s.t.} \quad & \mathcal{O} \cup \mathcal{A} \cup \mathcal{F}
\end{aligned}
\end{equation}

\subsection{A few guidelines and remarks on the method}
As mentioned in the above, the method enables to consider the carbon and financial cost of the system, although the user is free to leverage this feature or not (i.e., by manipulating the inputs of the problem). Similarly, the method considers the stochastic nature of the consumption, the irradiance and the carbon intensity of the grid. However, the user can run the problem with $\cs{N}{sc}{}=\cs{N}{tp}{}=1$ and perform a deterministic sizing of the resources. Note that in reality, the optimal sizing depends on the control scheme that will be applied to the system, as well as on the forecasting methods that will be used. To guarantee the coherence between the sizing of the resources and their operation, the user has to use the same forecasting methods and operational objectives adopted in the sizing stage.
\begin{rmk}
\label{rmk:weight}
    The weight $w$, in \SI{}{\g CO_2eq \per}¤, balances the trade-off between minimizing carbon emissions and minimizing costs. It represents the amount of \SI{}{\g CO_2eq} that need to be saved to justify a unit increase of the operational cost of the system. In \cref{eq:w interpretation}, the objective function (\ref{eq:objective function}) is divided by $w$ to provide its simpler and clearer interpretation. Indeed, the inverse of $w$ represents the financial cost of an emitted \SI{}{\g CO_2eq}. \hl{Therefore, users of the sizing tool can appropriately set $w$ according to preliminary studies on the value of a saved gram of carbon dioxide equivalents. The studies should consider the local policies (such as carbon taxes) and/or company-specific policies.}
    \begin{equation}
        \label{eq:w interpretation}
        \begin{aligned}
            \frac{1}{w}\cs{F}{obj}{}\left(\cs{P}{ess}{rated}, \cs{E}{ess}{rated}, \cs{P}{gen}{rated}\right)= \frac{1}{w}\left(\cs{C}{e}{pcc} + \cs{C}{e}{ess} + \cs{C}{e}{gen}\right) + \cs{c}{ess}{} + \cs{c}{gen}{} + \cs{c}{el}{energy} + \cs{c}{el}{power}
        \end{aligned}
    \end{equation}
\end{rmk}
\begin{rmk}
\label{rmk:Ppcc splitting}
$\cs{\textbf{P}}{pcc}{}$ is divided into $\cs{\textbf{P}}{pcc}{load},~\cs{\textbf{P}}{pcc}{gen}$ to enable different policies for the carbon and financial costs at the PCC, depending on whether the PCC is exporting or importing power. The application of the policies can be observed in the constraints for $\cs{C}{e}{pcc}$ and $\cs{c}{el}{energy}$ \sref{const:carbon pcc} and \sref{const:cost energy}. However, for the optimal solution of the problem to be relevant, $\cs{\textbf{P}}{pcc}{load},~\cs{\textbf{P}}{pcc}{gen}$ must be exclusive, meaning that at least one of them must be equal to zero at every timestep (i.e., power can not be imported and exported at the same time). \hl{This relation is guaranteed through constraints \sref{const:pcc split} to \sref{const:indicator pcc gen}, using the boolean variable $\cs{\textbf{z}}{pcc}{}$. In practice the complexity associated with a mixed-integer linear programming formulation is avoided using a continuous relaxation ($\cs{z}{pcc}{relaxed}$) of the indicator variable $\cs{z}{pcc}{}$ and the Big-M method \cite{cococcioni_big-m_2021}. In this case, \sref{const:indicator pcc load} and \sref{const:indicator pcc gen} are replaced by \sref{relaxation:pcc split}. It is important to note that inappropriate choices of the Big-M constants can lead to infeasible or inaccurate solutions and, therefore, the user should verify that the exclusivity of the variables is properly maintained. While a MILP formulation should be suitable to analyze few system sizings, it is too computationally demanding for the broader analysis required in this study (the analysis is not included here but resulted in a prohibitive computational effort).} 
Similarly, the battery power $\cs{\textbf{P}}{ess}{}$ is split into two components to account for system efficiency \sref{const:ess efficiency}. The exclusivity of the two variables $\cs{\textbf{P}}{ess}{charge}$ and $\cs{\textbf{P}}{ess}{discharge}$ is enforced and relaxed using the same approach.
\begin{subequations}
\label{relaxation:pcc split}
\begin{alignat}{1}
    &\cs{\textbf{P}}{pcc}{load} \leq \cs{M}{pcc}{} \cdot (1-\cs{z}{pcc}{relaxed})\\
    -&\cs{\textbf{P}}{pcc}{gen} \geq \cs{M}{pcc}{} \cdot \cs{z}{pcc}{relaxed}\\
    &0\leq \cs{z}{pcc}{relaxed} \leq 1
\end{alignat}
\end{subequations}
\end{rmk}

\begin{rmk}
    \label{rmk:p2e ratio}
    \hl{As already discussed before,} the parameter $\cs{r}{ess}{p2e}$ (defined in \cref{eq:rp2e}) is used to convexify the constraint \cref{const:cost ess}. This parameter should be selected carefully, as it reduces the set of possible ESS ratings to those with a fixed power-to-energy ratio. \hl{To get an idea of how this ratio impacts the sizing process (or to help the user to select a ratio), the sizing problem can be solved multiple times with different ratios. However, it is worth noting that, in reality, the parameter is limited to the options proposed by ESS manufacturers, which typically provide only a restricted range of power-to-energy ratios.}
\end{rmk}

\begin{rmk}
    \label{rmk:relaxation}
    Note that in many cases, some of the equality constraints in the problem might have to be relaxed (either for computational reasons or to make the problem feasible). For example, constraint \sref{const: ess as buffer} can be relaxed by replacing it by the two constraints $-\epsilon \leq \sum_{k=0}^{N-1}\left(\cs{\textbf{P}}{ess}{}\cdot\textbf{T}_\text{m}^\intercal\right)[k] \leq \epsilon$, where $\epsilon$ is a small-enough positive number chosen by the user. Similarly, constraint \sref{const: dispatch} can be relaxed by replacing it with constraints \sref{const: upper disp} and \sref{const: lower disp}, which results in setting a $\cs{\epsilon}{t}{}$ dispatch-tracking accuracy. Constraint \sref{const: avg disp} computes the dispatch plans as the expected value of the PCC power profile of each typical day. 
    \begin{subequations}
    \label{eq:dispatch tracking constraints}
        \begin{alignat}{1}
            &\cs{\textbf{P}}{pcc}{} \geq \cs{\textbf{P}}{pcc}{dispatch}\cdot\cs{\textbf{T}}{m}{}-\cs{\epsilon}{t}{}, \label{const: upper disp}\\
            &\cs{\textbf{P}}{pcc}{} \leq \cs{\textbf{P}}{pcc}{dispatch}\cdot\cs{\textbf{T}}{m}{}+\cs{\epsilon}{t}{}, \label{const: lower disp}\\
            &\cs{\textbf{P}}{pcc}{} \cdot \textbf{T}_{\text{m}}^{\intercal} = \cs{N}{sc}{}\cdot\cs{\textbf{P}}{pcc}{dispatch}, \label{const: avg disp}
        \end{alignat}
    \end{subequations}
\end{rmk}

\section{Case study: data and implementation context}
\label{case studies}
\subsection{Implementation}
The method presented in Section~\ref{method} is implemented using Jupyter Notebooks with \texttt{Python 3.11.8}. The \texttt{cvxpy} library \cite{diamond_cvxpy_nodate} is used to formulate the problem and license based \texttt{GUROBI} is used to solve it. License-free solvers (e.g., \texttt{CLARABEL}) that are natively supported by \texttt{cvxpy} can also be used, although they might be less efficient and/or produce less accurate solutions.

\subsection{Data availability}
This section provides a description of the data used in this research, with references. 

\subsubsection{Data center consumption}
\label{sec: dc consumption}
Publicly available data from \cite{wilkes_yet_2020} is used to model the consumption of the servers. The dataset contains a month of data (i.e., may 2019), with \SI{5}{\minute} granularity for ~50 \emph{Power Distribution Units} (PDUs), each PDU reports data in per units. The PDUs are aggregated in cells. Cell A is selected arbitrarily and the power attributed to production workloads \cite{sakalkar_data_2020} (i.e., the production\_power\_util field) of all the power distribution units in that cell (assuming every PDU in the cell has the same power rating) is aggregated. The aggregated data is then scaled up to a selected installed power.

\subsubsection{Grid carbon intensity} 
The grid carbon intensity data utilized in this study comes from a start-up specializing in solutions to monitor the cleanliness of electricity. More documentation can be found in \cite{emissium_emissium_nodate}. In particular, the start-up provides tools capable of tracking the Global Warming Potential over 100 years (GWP100) of a \SI{}{\kWh} with a \SI{15}{\minute} time granularity in different NUTS levels (Nomenclature of Territorial Units for Statistics). In this research, swiss cantonal data for the year 2023 is leveraged along with the data of german first level NUTS, enabling detailed analysis and insights into the temporal (and geographical) variations of carbon intensity in the german and swiss electricity grid.

\subsubsection{Electricity prices}
The day-ahead market prices come from the ENTSOE transparency platform, for year 2023 \cite{entsoe_day-ahead_2023}. Since day-ahead market prices are used, the injection and consumption tariffs are the same and the power price is not considered (i.e., is set to zero). Note that for different billing schemes (s.a. \cite{romande_energie_tarifs_2024}), the injection and consumption tariffs might be different and a power tariff might need to be incorporated. 

\subsubsection{Irradiance}
Historical data for the locations under study is recovered from the Copernicus Atmosphere Monitoring Service (CAMS). The data covers year 2023 and was accessed in August 2024 \cite{copernicus_atmosphere_monitoring_service_cams_cams_nodate, qu_fast_2017, schroedter-homscheidt_surface_2022}. 

\subsubsection{ESS and PV inputs}
The life-cycle footprint (GWP100) of the storage and photovoltaic system are recovered from the Ecoinvent database \cite{wernet_ecoinvent_2016}. In particular, the LCA values for the production of Li-ion LiMn2O4 batteries (reported in \cite{dominic_notter_battery_nodate}) and for the flat-roof installation of multi-Si PV systems in Switzerland (reported in \cite{niels_jungbluth_photovoltaic_nodate}) are used. The financial cost of a battery installation is recovered from \cite{yi_dispatch-aware_2023}, and the financial cost of a PV installation is recovered from \cite{christian_bauer_potentials_2017}. The study cases focus on battery storage systems because their high round-trip efficiency and their fast response time make them particularly well suited for short-term applications (such as day-ahead dispatching).

\subsection{Generation of scenarios}
Although forecasting stochastic variables is not the focal point of this paper, addressing scenario generation is essential to effectively study the method. In this paper, a combination of clustering and random sub-sampling (i.e., Monte Carlo simulations \cite{glasserman_monte_2010}) is used to generate scenarios. Specifically, seasonal clustering and weekday/weekend clustering are applied for the load, day-ahead prices and carbon emissions, as well as seasonal and energy-based clustering for the irradiance. Irradiance observations within a given season are grouped into three clusters based on their surface energy content (i.e., the integral of the irradiance over a day) using K-Means \cite{lloyd_least_1982}. The typical days to be considered by the sizing problem are selected so that they maintain the ratios between the populations of these clusters as accurately as possible. While respecting these ratios, a cluster is randomly assigned to every typical day. For the irradiance data, this step is valid under the assumption that information on the surface energy of the irradiance can be known day-ahead: it is reasonable to assume that information on the level of cloudiness of a day can be known 24\SI{}{\hour} ahead. \\
\hl{More accurate methods for scenario generation are likely to favor the penetration of photovoltaic and reduce the need for energy storage system capacity, since they are expected to lead to a smaller need for flexibility to achieve better day-ahead dispatchability.}

\subsection{Case study}
\label{sec:case study}
The case of a 1 \SI{}{\MW} rated data center in different locations across Switzerland and Germany is studied. Switzerland, where this research is based, provides a unique context due to its energy mix, while Germany serves as a relevant comparative case given its geographical proximity to Switzerland and its radically different energy mix. Sizings are performed over 21 typical days per season (3 per day of the week) with 20 scenarios per typical day (i.e., $\cs{N}{tp}{} = 84,~\cs{N}{sc}{}=20$), generating $\cs{M}{}{}=1680$ scenarios per stochastic variable. This choice is detailed in \ref{app:scenarios number}. Unless explicitly stated otherwise, a unitary power to energy ratio $\cs{r}{ess}{p2e} = 1$ is used. 

\section{Results}
\label{results}
\subsection{Single objective: carbon footprint reduction}
\label{sec: single objective}
\begin{figure}
        \centering
        \begin{subfigure}[b]{0.4\textwidth}
                \centering
                \includegraphics[width=\textwidth]{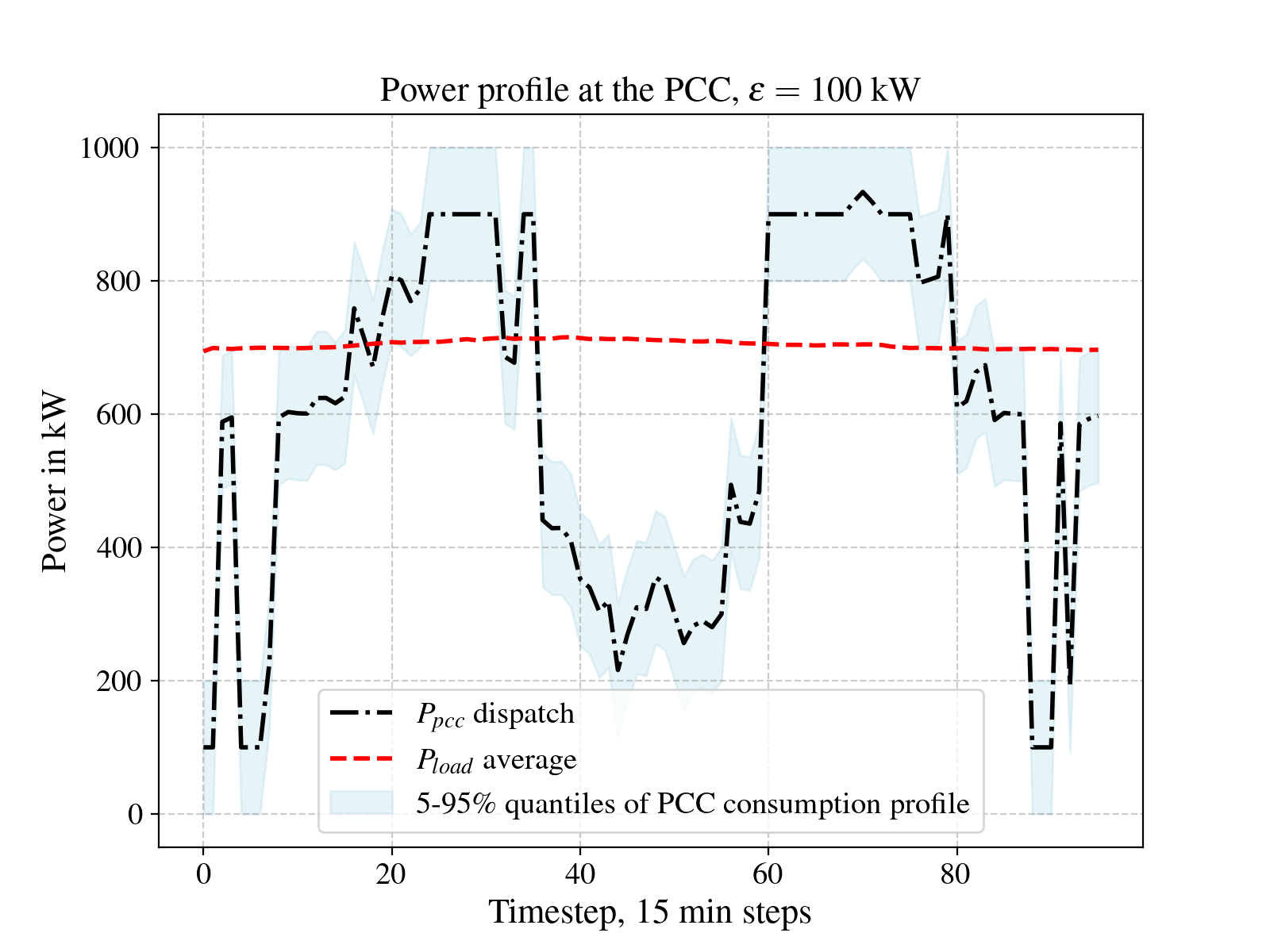}
                \caption{\hl{Dispatch and PCC power profiles.}}
                \label{fig:dispatch_zero_weight}
        \end{subfigure}%
        \begin{subfigure}[b]{0.4\textwidth}
                \centering
                \includegraphics[width=\textwidth]{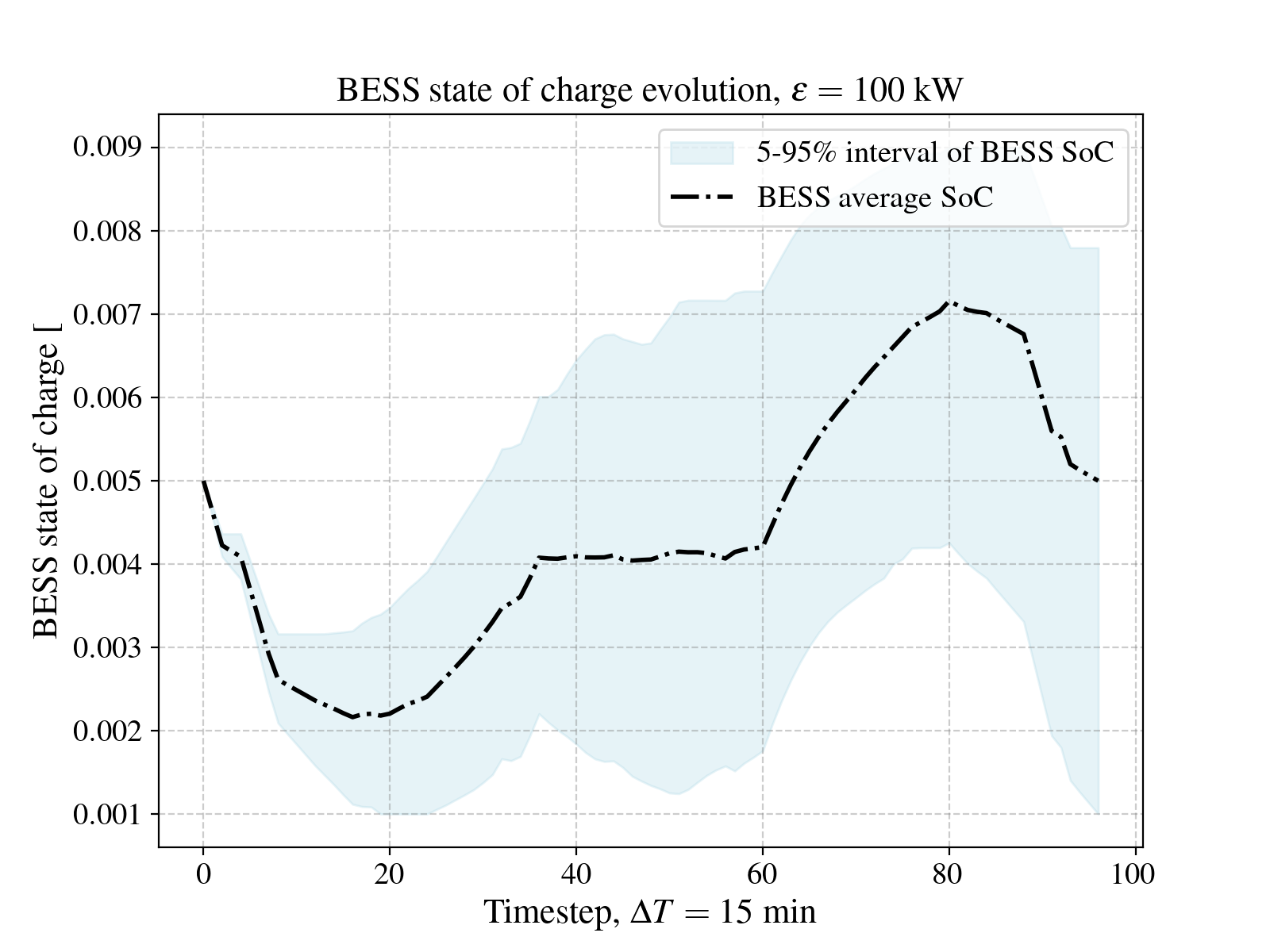}
                \caption{\hl{Evolution of the battery state of charge.}}
                \label{fig:soc_zero_weight}
        \end{subfigure}%

        ~ 
        \begin{subfigure}[b]{0.4\textwidth}
                \centering
                \includegraphics[width=\textwidth]{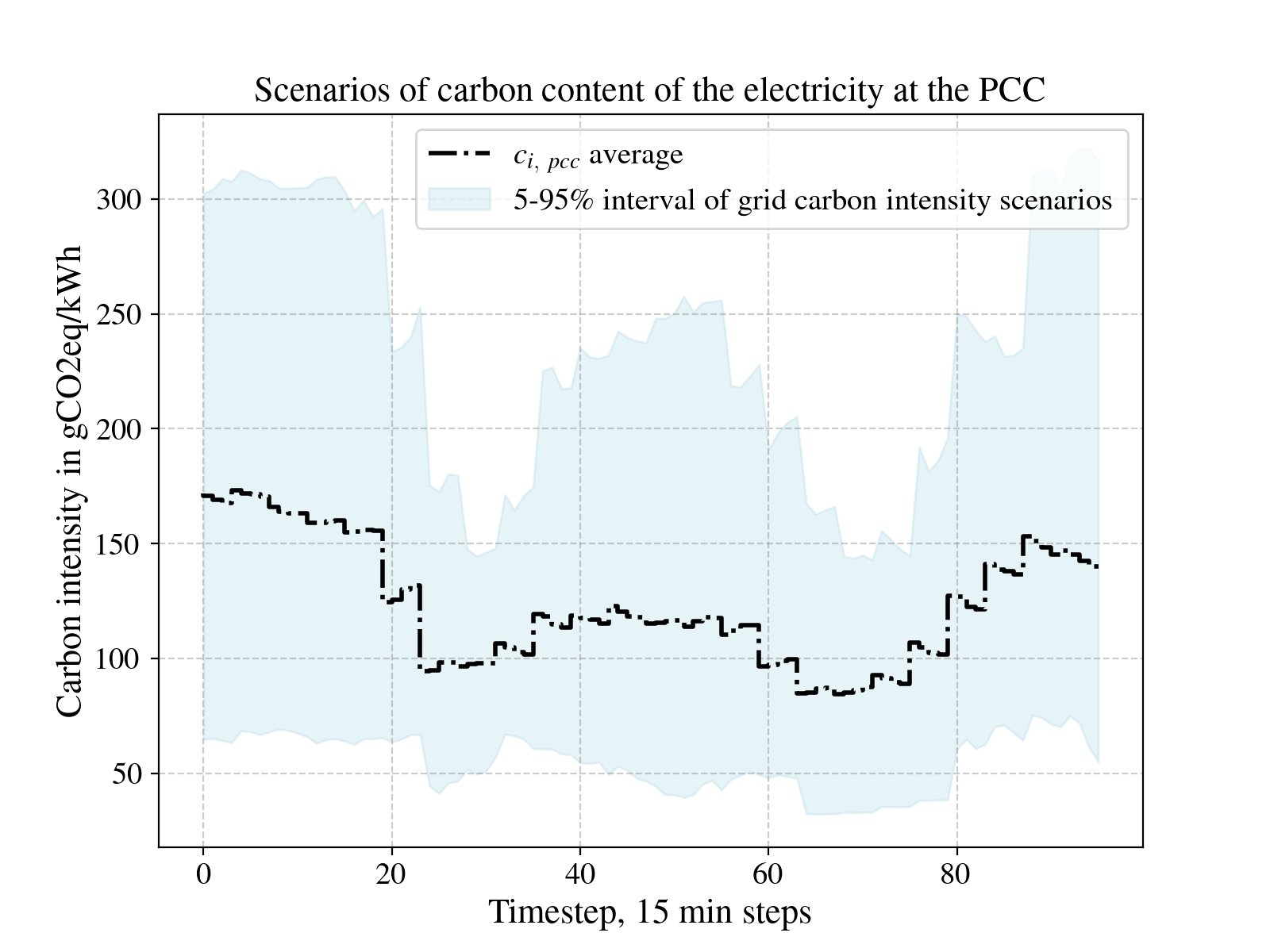}
                \caption{\hl{Average carbon intensity.}}
                \label{fig:ci_zero_weight}
        \end{subfigure}
        ~ 
        \begin{subfigure}[b]{0.4\textwidth}
                \centering
                \includegraphics[width=\textwidth]{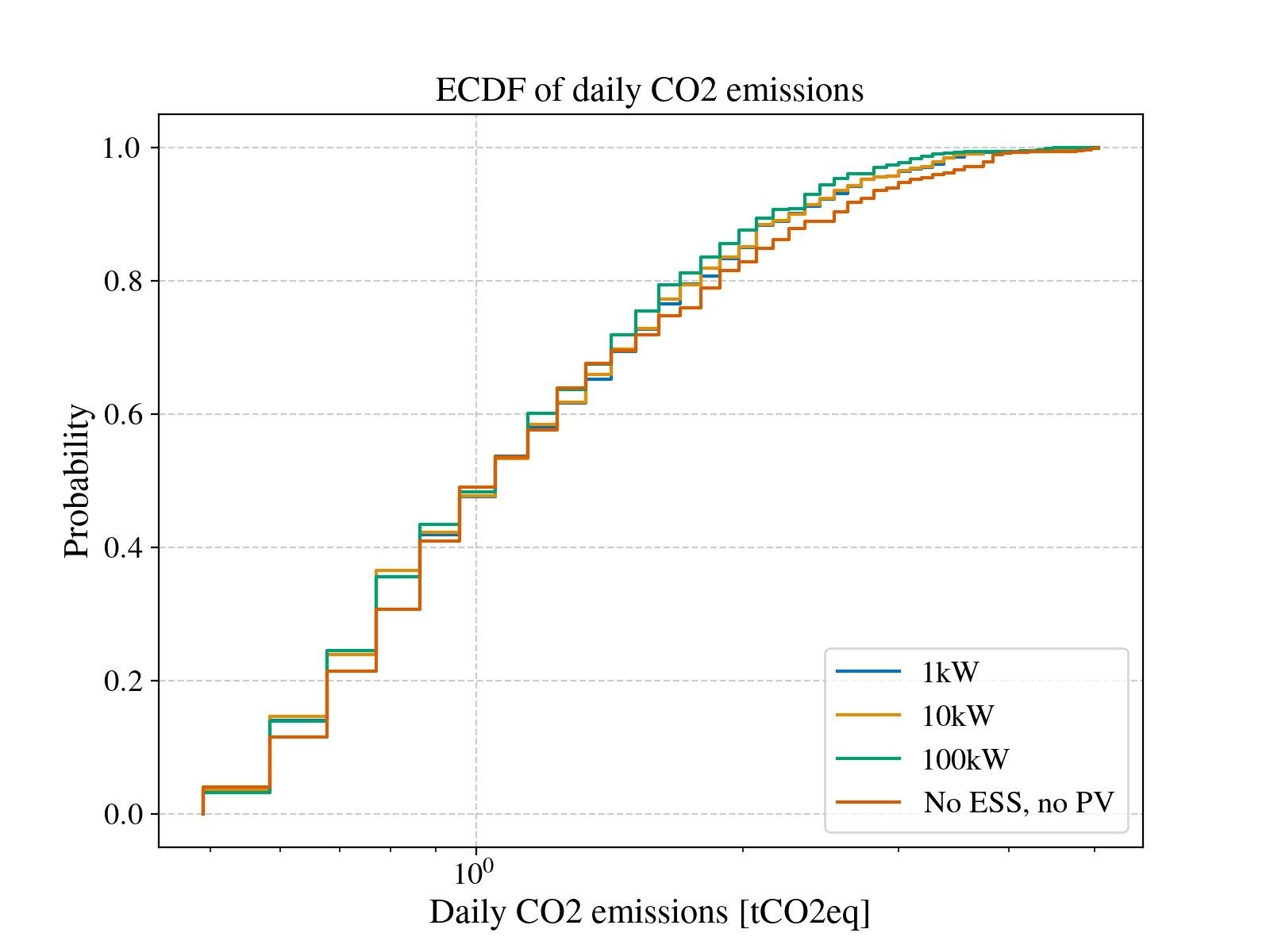}
                \caption{\hl{ECDF of the expected carbon emissions.}}
                \label{fig:ecdf_0}
        \end{subfigure}
        \caption{Detailed first typical day of a carbon-only sizing in canton Neuchâtel, Switzerland (NE).}\label{fig:w0}
\end{figure}

For the first analysis of the method, the weight $w$ is set to 0 (i.e., the financial aspect of the problem is disregarded), and the case of the Neuchâtel Swiss canton is studied\footnote{Canton Neuchâtel was selected as the case study in this section as it is a location with reasonable carbon reduction potential.}. Three levels of dispatch accuracy $\cs{\epsilon}{t}{}$ are studied (leveraging Remark~\ref{rmk:relaxation}): \SI{1}{\kW}, \SI{10}{\kW} and \SI{100}{\kW}, corresponding to a power tracking error of 0.1\%, 1\% and 10\% of the data center rated power. In \cref{fig:dispatch_zero_weight} and \cref{fig:soc_zero_weight}, scenarios for a random typical day of the sizing process are shown. \cref{fig:dispatch_zero_weight} shows the computed dispatch (the dash-dot black line) as well as the 5 and 95\% quantiles of the PCC power profiles and the average data center demand (the dashed red line), while \cref{fig:soc_zero_weight} shows the evolution of the battery state-of-charge. One can observe that the battery will discharge (i.e., the PCC power consumption is lower than the data center power demand) during the first hours of the day, because the carbon intensity is on a peak. Moreover, during the early morning hours, the battery starts to charge because the carbon content of the grid's electricity during those hours is lower (\cref{fig:ci_zero_weight}). Then, the PCC demand lowers to avoid the higher grid carbon intensity hours (and to self-consume solar production), before going back up during late afternoon hours (where the carbon intensity is lower). It is worth noting the spread in the battery state-of-charge at the end of the day since, depending on the scenario, the battery has to follow different trajectories. On average, as imposed by \cref{const: ess as buffer}, the battery acts like an energy buffer (as shown by the black dotted line, which shows the average SoC evolution).\\
The optimal capacity of the battery storage system is \SI{4.9}{\MW\hour}, \SI{4.5}{\MW\hour} and \SI{4.6}{\MW\hour} respectively. Since the power-to-energy ratio is equal to 1, the BESS power rating is \SI{4.9}{\MW}, \SI{4.5}{\MW} and \SI{4.6}{\MW}. For the PV system, the optimal power rating is \SI{1.1}{\MW}, \SI{1.1}{\MW} and \SI{1.3}{\MW}. These results appear very reasonable: the battery capacity is substantial because its carbon footprint is relatively low, allowing for carbon emission reductions via load shifting. Conversely, the PV system's capacity is comparable to the data center rated power, and slightly increases with $\cs{\epsilon}{t}{}$, as lower tracking accuracy allows for more stochasticity in the data center complex. Additionally, the battery rated capacity decreases when $\cs{\epsilon}{t}{}$ increases as the battery is the only asset able to compensate the stochasticity in the system and, therefore, allowing for less precise tracking requires less BESS installation. It is worth noting that the choice of $\cs{\epsilon}{t}{}$ is user-specific, as it allows the user to decide the importance of having a dispatchable system (i.e., large enough $\cs{\epsilon}{t}{}$ will completely disregard the objective of achieving dispatchability). In the remainder of the paper, dispatchability is considered as a crucial aspect of the data center daily operations, given the current costs associated to imbalance prices in electricity. Therefore, a tracking accuracy of 0.1\% is used.\\
In \cref{fig:ecdf_0}, the \textit{Empirical Cumulative Distribution Function} (ECDF) of the carbon emissions demonstrates that the method decreases the system's \textit{Carbon Emissions} (CE). The main reduction comes from high emission days, which are significantly reduced by the sizing process. \cref{tab:w0 sizing} shows that the sizing method reduces both the average daily emissions as well as their standard deviation. 
\begin{table}[h]
    \centering
    \begin{tabular}{c|c|c|c|c}
         Daily emissions                    & No ESS, no PV & $\cs{\epsilon}{t}{} = \SI{1}{\kW}$ & $\cs{\epsilon}{t}{} = \SI{10}{\kW}$ & $\cs{\epsilon}{t}{} = \SI{100}{\kW}$\\ \hline
         Average in \SI{}{tCO_2eq}       & 1.42 & 1.36 & 1.35 & 1.3 \\ 
         Standard dev. in \SI{}{tCO_2eq} & 0.87 & 0.75 & 0.75 & 0.68 \\ 
    \end{tabular}
    \caption{Metrics for daily carbon emissions, sizing with $w=0$.}
    \label{tab:w0 sizing}
\end{table}

\subsection{Pareto front}
\label{sec:pareto fronts}
In this section, the sensitivity of the sizing process to the weight $w$ is studied, for three Swiss cantons (i.e., Vaud-VD, Aargau-AG and Neuchâtel-NE) and two NUTS1 areas of Germany (i.e., Bayern-DE2 and Schleswig-Holstein-DEF). Aargau and Neuchâtel are selected as they are the Swiss cantons with lowest and highest average grid carbon intensity respectively (as shown in \cref{fig:ci_cantonal}), while the average of canton Vaud is close to the Swiss average grid carbon intensity. Bayern and Schleswig-Holstein are selected to represent the northern and southern part of Germany. In this context, the evolution of the ESS and PV ratings, as well as the evolution of the carbon and financial objectives are shown in \cref{fig:pareto_front}. 

\subsubsection{Evolution of the objectives}
By focusing on the plots on the right side of \cref{fig:pareto_front}, the impact of the weight (i.e., $w$) on the sizing objectives can be studied. As a reminder, the objectives are the reduction of the system's carbon footprint and of the system's costs, as detailed in \cref{eq:objective function}. As expected, increasing the weight tends to increase carbon emissions while reducing costs. This happens because larger weights indicate that the DCO puts the emphasis on the economic aspect of the system operation, therefore prioritizing cost savings over carbon savings. In \cref{fig:pareto_front}, the black dotted lines show the expected daily objectives for the base case (i.e., for the operation of the data center alone; without PV, without BESS and without demand side control). The base case expected carbon emissions for the different regions are significantly different as the historical carbon intensity of the electricity consumed in each region varies, and the carbon saving opportunities strongly depend on the base case carbon emissions. In fact, the maximum expected footprint reduction (i.e., the reduction when $w=0$) is approximately 49.6\% for a sizing in the Bayern region, 14.7\% in Schleswig-Holstein, 4\% in canton Neuchâtel, 0.3\% for canton Vaud and -2.6\% for canton Aargau. It is interesting to observe that for a region in which the carbon intensity of the consumed electricity is very low (such as Aargau), the underlying constraint of achieving the dispatchability of the system (i.e. being able to track a day-ahead consumption plan) increases the expected carbon emissions of the system, even if the sizing is performed with the carbon footprint objective only. In terms of costs, sizing the DERs using low weights can increase the operational costs significantly (e.g., the $w=0$ sizing of canton Neuchâtel leads to a 21.4\% increase and to 118.3\% in Bayern). Larger weights still lead to 2-3\% cost increases compared to the base case scenario for all regions, meaning that achieving dispatchability tends to increase the costs of the system operation. Moreover, lowering the costs increase the carbon emissions of the system by up to 6.7\% for canton Aargau, 1\% for Vaud and 0.3\% for Neuchâtel. For the case of Germany, carbon emissions are reduced even at higher weights, by 0.1\% in Schleswig-Holstein and 4.1\% in Bayern.\\
It is important to remind that, as discussed in \cref{sec:rpe analysis}, this section presents the results for sizings with a BESS power-to-energy ratio equal to 1. For a fixed weight and a fixed location, this ratio might not be optimal: emissions and costs could be further reduced by selecting the optimal ratio for each sizing instance. In this section, however, the focus is on analyzing the sensitivity of the system to the weight only.  
\subsubsection{Evolution of the decision variables}
\begin{figure}
    \centering
    \includegraphics[width=.65\textwidth]{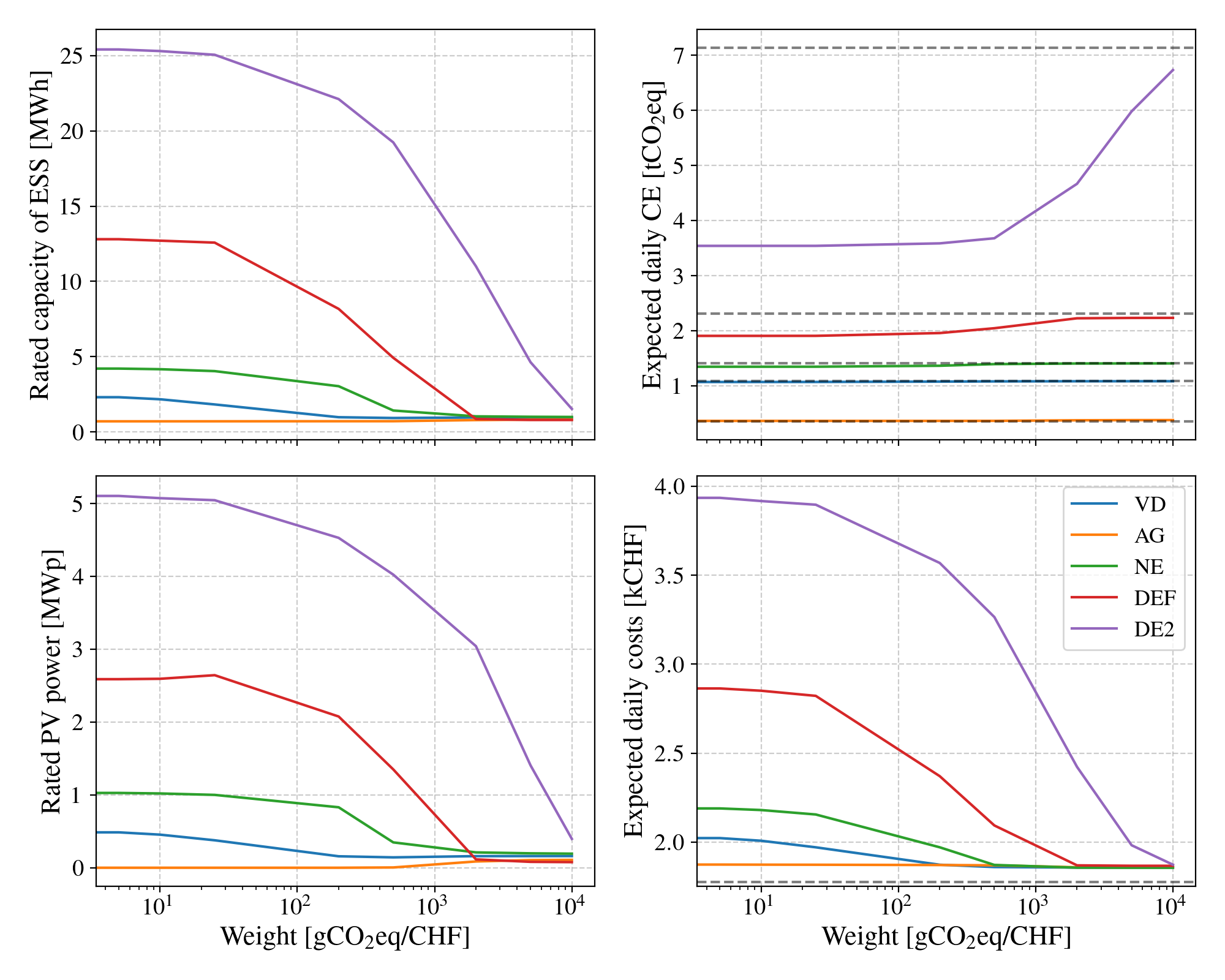}
    \caption{\hl{Evolution of decision variables and objectives with increasing $w$, for three Swiss cantons and two NUTS1 regions in Germany.}}
    \label{fig:pareto_front}
\end{figure}

The plots on the left side of \cref{fig:pareto_front} show how \hl{the rated capacity of BESS} and the power rating of photovoltaics evolve with increasing weight $w$. The general trend for the PV rated power is that it decreases with the weight, meaning that it tends to increase the operational costs of the systems. In Germany, since the grid's carbon intensity is much larger than that of Switzerland (i.e. there is an average factor of 6 between VD and DE2), low weights require the most installation of PV. Thus, the amount of installed PV generation for low weights depends on the grid carbon intensity at the location: if it is lower than the PV generation carbon intensity (e.g., canton AG), little to no PV is installed, because it would paradoxically increase the carbon footprint of the system. If it is larger (e.g., NUTS DE2), more PV is installed to reduce the footprint. For larger weights, the amount of installed PV is similar in all locations. The general trend for the rated capacity of the BESS follows that of the PV rated power. Indeed, with increased PV power, the stochasticity of the system increases, thus requiring a larger battery capacity to achieve dispatchability. Moreover, for low weights, and as discussed in the previous sub-section, the battery can help with the carbon emissions reduction. This observation implies that the battery can shift the electricity consumption of the system to periods where the grid carbon intensity is low, but that it is an expensive process because of the battery cost (since the rated capacity drops significantly between $w=0$ and $w=1000$ \SI{}{\g CO_2eq \per CHF}). \\
The optimal size of the resources is highly dependent on the location of the data center complex. For instance, sizings in canton Neuchâtel may suggest to install up to 2 times more battery capacity than sizings in canton Vaud, and up to 6 times more battery capacity than sizings in canton Aargau. Sizings in Bayern suggest battery capacities up to 6 times larger than those of Neuchâtel. Moreover, while the sizing process (with $w=0$) suggests to install over \SI{5}{\MW} of PV generation in Bayern, it suggests to only install \SI{500}{\kW} in canton Vaud and to not install PV in Aargau. Larger weights reduce the dispersion of the ratings of the resources, as the day-ahead market electricity prices are considered to be the same for every region, and less importance is given to the carbon costs. 
\subsection{Sensitivity to the BESS power-to-energy ratio}
\label{sec:rpe analysis}
\begin{figure}
    \centering
    \includegraphics[width=.7\textwidth]{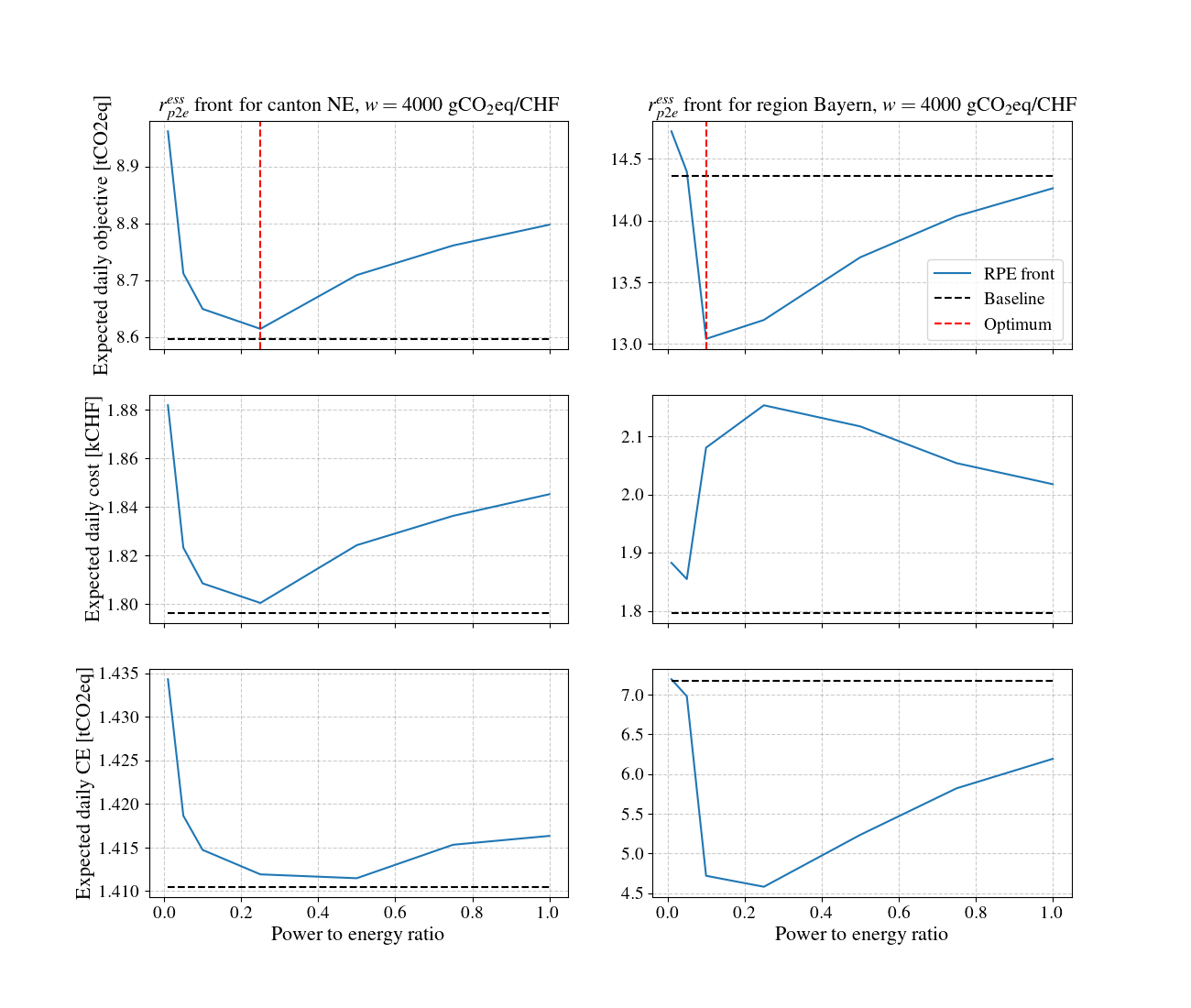}
    \caption{\hl{Evolution of objectives with increasing power-to-energy ratio, for two regions.}}
    \label{fig:rpe_front}
\end{figure}

As detailed in \cref{Formulation}, the power-to-energy ratio of the BESS $\cs{r}{ess}{p2e}$ was introduced as an input parameter to convexify the sizing problem formulation. In the results presented above, the parameter is fixed to $\cs{r}{ess}{p2e}=1$ (i.e., at 100\% state of charge, the battery can provide $\cs{P}{ess}{rated}$ for \SI{1}{\hour}), which may not be the optimal ratio for some sizing problems. As discussed in Remark \ref{rmk:p2e ratio}, the optimal ratio can be identified by performing multiple sizings with different values of $\cs{r}{ess}{p2e}$ and finding the ratio that leads to the smallest value of the objective function. In this section, this process is performed for the case of sizings in canton Neuchâtel and in the Bayern region, with $w=4000$\SI{}{\g CO_2eq \per CHF}. \\ 
Figure \ref{fig:rpe_front} shows the evolution of the objectives as the power-to-energy ratio increases from 0.01 to 1, for canton Neuchâtel on the left and for the Bayern region on the right. The black dotted lines show the baseline expected costs for each objective, while the red dotted lines show the optimal ratios. The top graphs show the evolution of the overall objective, while the middle and bottom graphs show the evolution of the daily cost and footprint respectively. From the top graphs of both regions, $\cs{r}{ess}{p2e} = 0.25$ and $\cs{r}{ess}{p2e} = 0.1$ (i.e., as supported by the vertical red dotted lines) can be selected as the optimal ratios for the proposed sizings. Note that the location of the data center impacts the optimal power-to-energy ratio, as well as the evolution of the sub-objectives (e.g., the middle graphs are evolving in almost opposing ways). Also note that the middle left graph (i.e., the evolution of the operational costs in canton Neuchâtel) shows that a properly sized battery is expected to reduce the expected operational costs of the system compared to the naive selection of $\cs{r}{ess}{p2e}=1$.

\subsection{Discussion on the results}
The results section used the proposed sizing tool to optimize BESS and PV systems coupled with data centers located in Switzerland and Germany. In Switzerland, while the tool achieves a maximum reduction in carbon emissions of approximately 4\%, this comes with a substantial increase in system costs, up to 21\%. Although these results show a marginal impact, this is mainly a consequence of the low carbon intensity of the Swiss electricity mix. Indeed, in Germany, the tool achieves maximum carbon reductions of up to 49.6\%, though this results in a doubling of operational costs. The analysis highlights that the optimal sizing of BESS and PV systems to achieve data center dispatchability is highly dependent on the data center location. For example, despite the relatively short distance of only \SI{109}{km} between the city of Neuchâtel and the city of Aarau, the sizing results for their cantons differ significantly. This demonstrates the necessity to use geographically accurate data when leveraging DER for cost and carbon reduction and the proposed framework can be effectively used to address these location-specific challenges. It is reasonable to suppose that case studies in other countries would lead to different results as, according to \cite{electricitymaps_electricity_nodate} the 2023 yearly average carbon intensity of Switzerland was \SI{86}{gCO_2eq/kWh}, while it was 4.5 times larger in the United States, Germany, and Ireland (i.e., approximately \SI{400}{gCO_2eq/kWh}), and 7.7 times larger in India (i.e., approximately \SI{660}{gCO_2eq/kWh}). Sizing BESS and PV for dispatchable data centers in countries with higher grid carbon intensity (s.a. the U.S., India, etc.) would likely lead to even more opportunities in terms of carbon footprint reduction. Finally, this results section assumes that achieving the dispatchability of the data center complex is an objective of the DCO. It is worth noting that this service does not only benefit the grid (by removing the stochastic behaviour of the power consumption profile), but would likely decrease the intra-day operational costs of the system, as the DCO would not have any imbalance fees to account for. 

\section{Conclusions}
\label{conclusion}
This work proposes a carbon and cost-aware framework to size energy storage systems and photovoltaic generation in the context of a data center aiming at achieving dispatchability, and presents an analysis of the framework in the Swiss and German contexts. The tool can be leveraged by data center operators to easily design a sizing process that takes into account their particular context and needs. Custom grid carbon intensities, DER life cycle assessments, GHI data, load data and custom forecasting methods can be jointly used with the framework. Moreover, user specific inputs provide some flexibility to the users (e.g., the tracking accuracy, the weight $w$, the power to energy ratio, etc.). \\
The analysis of the method in the Swiss and German contexts highlights the relevance of using geographically granular data in the sizing process, and shows that the method can be successfully used in that perspective. A large difference in the optimal sets of resources is observed: sizings in Bayern suggest BESS capacity ratings up to 36 times larger than sizings in Aargau. Achieving the dispatchability of a data center complex in Bayern is expected to decrease its carbon emissions by approximately 49.6\% at best, while, for a DC in Aargau, it is expected to increase its emissions by 2.6\% at least. In terms of costs, achieving dispatchability tends to increase the operational costs in all considered regions, although the careful selection of the power-to-energy ratio of the ESS can reduce the impact on operational costs (e.g. in Neuchâtel and with $w=4000$\SI{}{gCO_2 eq/CHF}, the optimal ratio reduces the operational costs by approximately 3\% compared to the naive selection where a unitary ratio is selected).\\
The main limitation of this work is that the cost of power imbalances are neglected and, therefore, further potential cost savings are not considered. \hl{It might also be relevant to study the impact of PV curtailment, as it would increase the flexibility of the system and could result in lower needs for energy storage. In future work, these limitations will be addressed, as the focus will shift towards the day-ahead and intra-day operation of data centers with co-located energy storage and curtailable photovoltaic generation, and the economic impact of imbalances will be thoroughly studied.}
\newpage
\section*{Author contributions: CRediT}
\textbf{Mario Paolone}: Supervision, conceptualization, funding acquisition, resources, validation, visualization, writing - review and editing \\
\textbf{Enea Figini}: Investigation, methodology, conceptualization, validation, visualization, software, formal analysis, data curation, writing - original draft.
\section*{Funding}
This work is supported by École polytechnique fédérale de Lausanne, through the Heating Bits Solutions 4 Sustainability project. 
\section*{Declaration of competing interest}
The authors declare that they have no known competing financial
interests or personal relationships that could have appeared to influence
the work reported in this paper.

\newpage
\label{nomenclature}
\nomenclature[01]{$\cs{N}{tp}{}$}{The number of typical periods considered in the optimization process}
\nomenclature[02]{$\cs{N}{sc}{}$}{The number of scenarios per typical time horizon}
\nomenclature[03]{$\cs{N}{}{}$}{The number of steps in a typical time horizon}
\nomenclature[031]{$\cs{M}{}{}$}{The total number of scenarios ($M=\cs{N}{tp}{}\cdot\cs{N}{sc}{}$)}
\nomenclature[0301]{$\cs{\Delta T}{}{}$}{The duration in hours of a discrete time step}
\nomenclature[021]{$\cs{W}{}{}$}{The duration in hours of the considered time horizon}
\nomenclature[022]{$\cs{W}{days}{}$}{The duration in days of the considered time horizon}
\nomenclature[07]{$\cs{\textbf{P}}{ess}{}$}{The ESS power in \SI{}{\kW}}
\nomenclature[08]{$\cs{E}{ess}{rated}$}{The rated energy of the ESS in \SI{}{\kWh}}
\nomenclature[09]{$\cs{P}{gen}{rated}$}{The rated power of the PV generation plant in \SI{}{\kW}}
\nomenclature[10]{$\cs{\textbf{E}}{ess}{}$}{The ESS energy in \SI{}{\kWh}}
\nomenclature[11]{$\cs{\textbf{P}}{pcc}{}$}{The power at the Point of Common Coupling (the power at the grid connection point) in \SI{}{\kW}}
\nomenclature[12]{$\cs{\textbf{P}}{pcc}{load}$}{The imported power at the PCC in \SI{}{\kW}}
\nomenclature[13]{$\cs{\textbf{P}}{pcc}{gen}$}{The exported power at the PCC in \SI{}{\kW}}
\nomenclature[1301]{$\cs{P}{pcc}{rated}$}{The rated power at the PCC (i.e., maximum power that can flow at the PCC)}
\nomenclature[131]{$\cs{\textbf{z}}{pcc}{}$}{Indicator variable used to determine the sign of the power at the PCC in the MILP formulation}
\nomenclature[1311]{$\cs{\textbf{z}}{pcc}{relaxed}$}{The continuous indicator variable used to determine the sign of the power at the PCC in the relaxed formulation}
\nomenclature[132]{$\cs{M}{pcc}{}$}{A large-enough constant used to guarantee the exclusivity of $\cs{P}{pcc}{load}$ and $\cs{P}{pcc}{gen}$ in the relaxed formulation}
\nomenclature[14]{$\cs{\textbf{P}}{gen}{}$}{The power generated by the PV plant in \SI{}{\kW}}
\nomenclature[15]{$\cs{\textbf{P}}{ess}{conv}$}{The power of the ESS, on the grid side in \SI{}{\kW}}
\nomenclature[16]{$\cs{\textbf{P}}{ess}{charge}$}{The charging power of the ESS, on the storage side in \SI{}{\kW}}
\nomenclature[17]{$\cs{\textbf{P}}{ess}{discharge}$}{The discharging power of the ESS, on the storage side in \SI{}{\kW}}
\nomenclature[171]{$\cs{\textbf{z}}{ess}{}$}{Indicator variable used to determine the sign of the power at the ESS in the MILP formulation}
\nomenclature[1711]{$\cs{\textbf{z}}{ess}{relaxed}$}{The continuous indicator variable used to determine the sign of the power at the ESS in the relaxed formulation}
\nomenclature[172]{$\cs{M}{ess}{}$}{A large-enough constant used to guarantee the exclusivity of $\cs{P}{ess}{charge}$ and $\cs{P}{ess}{discharge}$ in the relaxed formulation}
\nomenclature[18]{$\cs{\textbf{P}}{pcc}{dispatch}$}{The dispatches of the PCC power in \SI{}{\kW}}
\nomenclature[181]{$\cs{\textbf{T}}{m}{}$}{Transformation matrix to duplicate columns of a matrix $\cs{N}{sc}{}$ times}
\nomenclature[19]{$\cs{\textbf{P}}{pcc}{max}$}{The maximum power of each typical period in \SI{}{\kW}}
\nomenclature[20]{$\cs{P}{ess}{rated}$}{The rated power of the ESS in \SI{}{\kW}}
\nomenclature[201]{$\cs{P}{ess}{rated, max}$}{The upper bound for ESS rated power candidates in \SI{}{\kW}}
\nomenclature[21]{$\cs{E}{ess}{start}$}{The energy to be contained in the ESS at the start of each typical period in \SI{}{\kWh}}
\nomenclature[22]{$\cs{E}{ess}{min}$}{The minimum energy allowed in the ESS in \SI{}{\kWh}}
\nomenclature[23]{$\cs{E}{ess}{max}$}{The maximum energy allowed in the ESS in \SI{}{\kWh}}
\nomenclature[24]{$\cs{C}{e}{pcc}$}{Equivalent carbon emissions of the energy imported from the grid over a given time window in \SI{}{\gram CO_2eq}}
\nomenclature[25]{$\cs{C}{e}{ess}$}{Equivalent carbon emissions of the ESS over a given time window in \SI{}{\gram CO_2eq}}
\nomenclature[26]{$\cs{C}{e}{gen}$}{Equivalent carbon emissions of the PV plant over a given time window in \SI{}{\gram CO_2eq}}
\nomenclature[27]{$\cs{c}{ess}{}$}{Equivalent cost of the ESS over a given time window in ¤}
\nomenclature[28]{$\cs{c}{gen}{}$}{Equivalent cost of the PV plant over a given time window in ¤}
\nomenclature[29]{$\cs{c}{el}{energy}$}{Equivalent cost of energy imports over a given time window in ¤}
\nomenclature[30]{$\cs{c}{el}{power}$}{Equivalent cost of maximum imported power over a given time window in ¤}
\nomenclature[31]{$\cs{\textbf{P}}{load}{}$}{The load power consumption in \SI{}{\kW}}
\nomenclature[32]{$\cs{\textbf{i}}{ghi}{}$}{The global horizontal irradiance in \SI{}{\W\per\m^2}}
\nomenclature[33]{$\cs{\textbf{C}}{i}{pcc}$}{Carbon intensity of the imported electricity in \SI{}{\gram CO_2eq \per\kWh}}
\nomenclature[34]{$\cs{\textbf{p}}{el}{cons}$}{The price of energy consumption in ¤/\SI{}{\kWh}}
\nomenclature[35]{$\cs{\textbf{p}}{el}{inj}$}{The price of energy injection in ¤/\SI{}{\kWh}}
\nomenclature[36]{$w$}{The weight of the economical part of the objective function in \SI{}{\g CO_2eq\per}¤}
\nomenclature[37]{$\cs{P}{pcc}{rated}$}{The power rating of the grid connection point in \SI{}{\kW}}
\nomenclature[38]{$\cs{C}{i}{ess}$}{Equivalent carbon intensity of the ESS in \SI{}{\gram CO_2eq \per\kWh}}
\nomenclature[39]{$\cs{C}{e, LCA}{ess}$}{Equivalent carbon emissions needed to manufacture a \SI{}{\kWh} of ESS in \SI{}{\gram CO_2eq \per\kWh}}
\nomenclature[391]{$\cs{C}{e, LCA}{gen}$}{Equivalent carbon emissions needed to manufacture a \SI{}{\kW} of PV plant in \SI{}{\gram CO_2eq \per\kW}}
\nomenclature[392]{$\cs{a}{ess}{}$}{Amount of aging of the ESS}
\nomenclature[393]{$\cs{c}{ess}{life}$}{Cost of the ESS over its lifetime}
\nomenclature[40]{$\cs{r}{ess}{p2e}$}{The power-to-energy ratio of the ESS}
\nomenclature[41]{$\cs{r}{gen}{ghi2p}$}{User defined constant to adjust the slope of the ghi to power model of the PV plant}
\nomenclature[42]{$\cs{i}{ghi}{max}$}{The maximum GHI at the location in \SI{}{\W \per \m^2}}
\nomenclature[43]{$\cs{SoC}{ess}{min}$}{The minimum allowed state of charge of the ESS}
\nomenclature[44]{$\cs{SoC}{ess}{max}$}{The maximum allowed state of charge of the ESS}
\nomenclature[45]{$\cs{SoC}{ess}{start}$}{The state of charge of the ESS at the start of every typical period}
\nomenclature[46]{$\cs{L}{ess}{cycles}$}{Expected lifetime of the ESS in cycles}
\nomenclature[47]{$\cs{L}{ess}{calendar}$}{Expected lifetime of the ESS in hours}
\nomenclature[48]{$\cs{L}{gen}{calendar}$}{Expected lifetime of the PV plant in hours}
\nomenclature[49]{$\cs{c}{ess}{energy}$}{Overall cost of a \SI{}{\kWh} of installed battery storage in ¤/\SI{}{\kWh}}
\nomenclature[50]{$\cs{c}{ess}{power}$}{Overall cost of a \SI{}{\kW} of installed battery storage in ¤/\SI{}{\kW}}
\nomenclature[51]{$\cs{c}{gen}{power}$}{Overall cost of a \SI{}{\kW} of installed PV generation in ¤/\SI{}{\kW}}
\nomenclature[52]{$\cs{p}{el}{power}$}{Cost per \SI{}{\kW} of the daily maximum power consumed at the PCC in ¤/\SI{}{\kW}}
\nomenclature[53]{$\cs{\eta}{ess}{}$}{Efficiency of the energy storage system}
\nomenclature[54]{$\cs{\epsilon}{t}{}$}{The dispatch tracking accuracy in \SI{}{\kW}}
\printnomenclature

\appendix
\section{Choice of the number of scenarios}
\label{app:scenarios number}
\begin{figure}[h]
  \centering
  \includegraphics[width=0.6\linewidth]{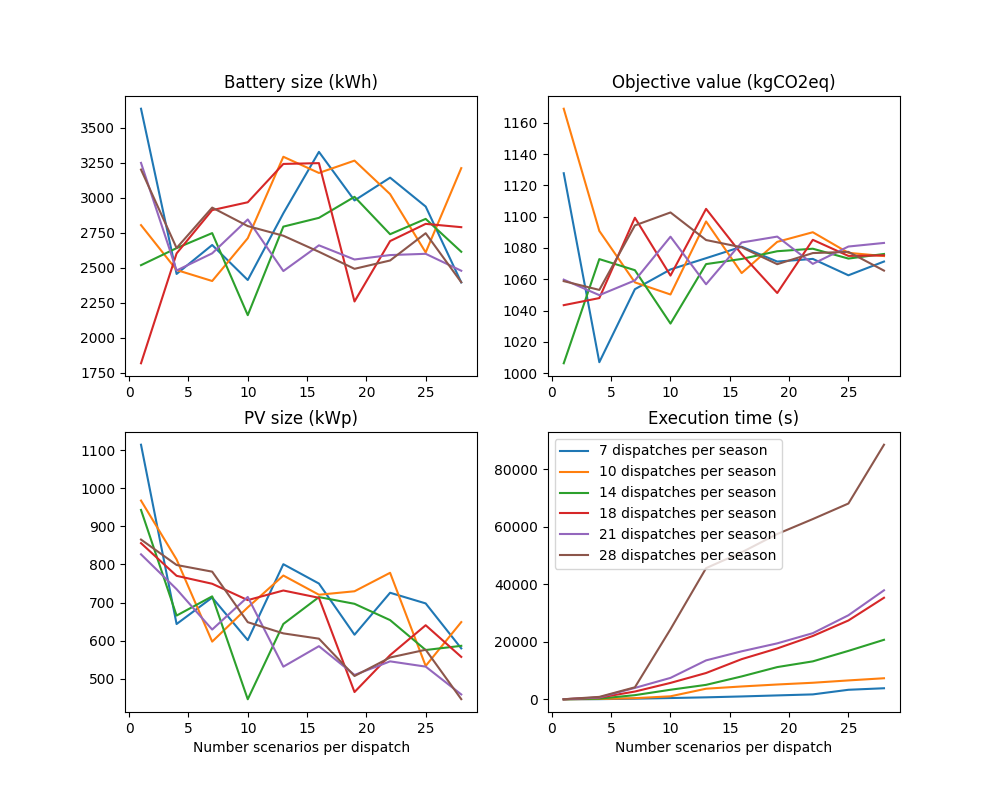}
  \caption{Evolution of sizing results for different sets of $\cs{N}{td}{}$ and $\cs{N}{sc}{}$.}
  \label{fig:scenario gen params}
\end{figure}

There are two variables to be selected: $\cs{N}{tp}{}$ and $\cs{N}{sc}{}$. In the selection process, the number of typical days in a season ($\cs{N}{td}{}$) is introduced. Note that $\cs{N}{tp}{}=4\cs{N}{td}{}$. To select them, the combinations of $\cs{N}{td}{}$ and $\cs{N}{sc}{}$ are progressively increased and the proposed sizing algorithm executed for each set. The results are shown in \cref{fig:scenario gen params}. The most relevant observation is that the objective value tends to a constant of approx. 1075 \SI{}{\kg CO_2eq} for $\cs{N}{sc}{}\geq20$. The left figures show the evolution of the decision variables, and demonstrate that for $\cs{N}{td}{}\geq21$, the optimal values of the decision variables are very close (in particular for values of $\cs{N}{sc}{}\geq20$). For the analysis purposes of this paper, the combination $\cs{N}{td}{}=21$ and $\cs{N}{sc}{}=20$ is selected, as it provides a good compromise between the insensitivity of the optimal results with respect to the numbers of scenarios vs. the execution times. This simple analysis is suggested to be adopted by users interested in implementing the proposed sizing method. 

\newpage
\section{Carbon intensities across Switzerland}
\begin{figure}[h]
    \centering
    \includegraphics[width=.45\textwidth]{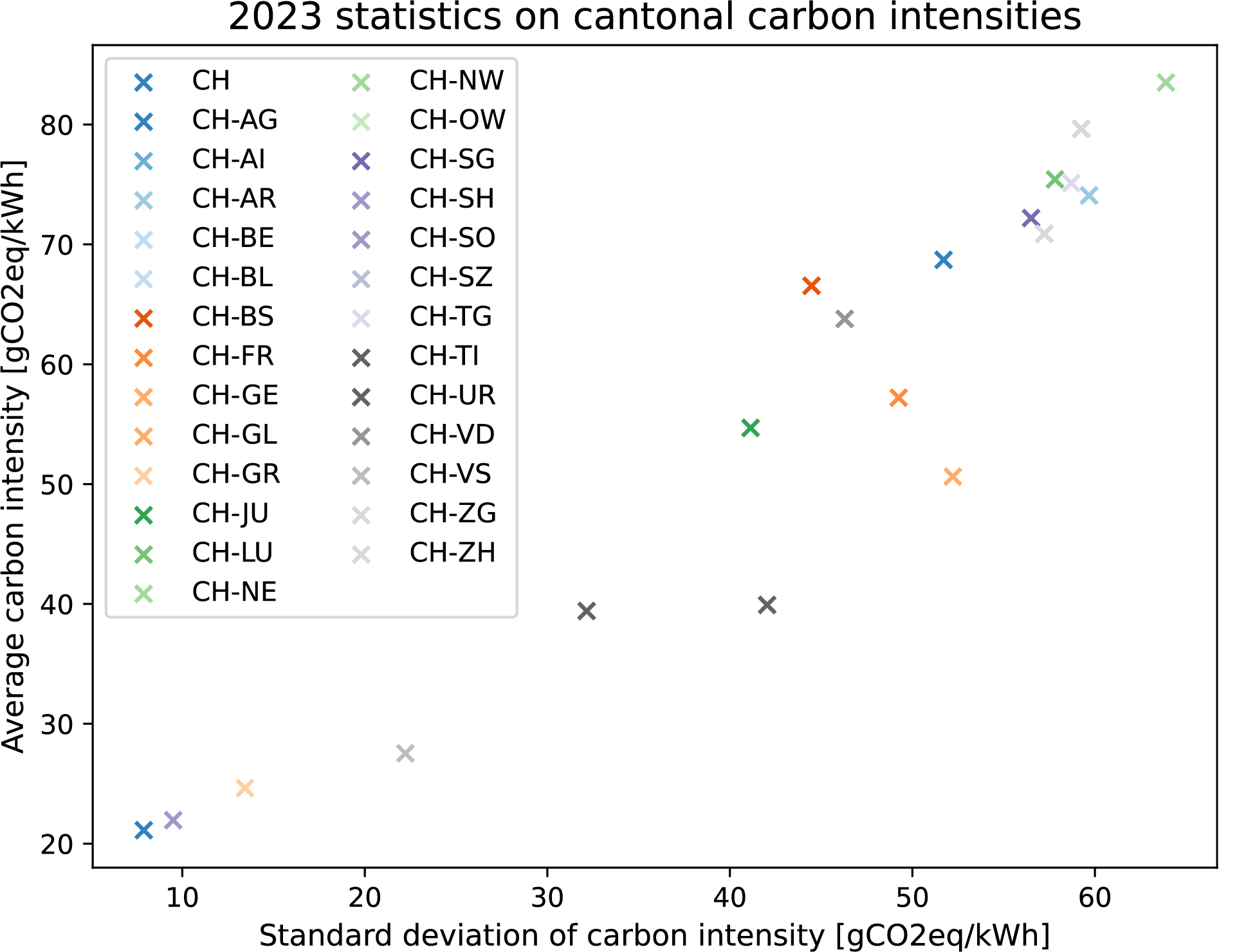}
    \caption{Average and standard deviation of the grid carbon intensities in Switzerland.}
    \label{fig:ci_cantonal}
\end{figure}

\cref{fig:ci_cantonal} shows the 2023 average carbon intensity of all Swiss cantons. Cantons AG and NE are selected as the extreme cases, while VD is selected to represent the average Swiss case. 
\newpage



\bibliographystyle{elsarticle-num}
\bibliography{references}







\end{document}